\documentclass[12pt,a4paper]{article}

\pdfoutput=1

\usepackage[utf8x]{inputenc}
\usepackage[T1]{fontenc}
\usepackage{lmodern}
\usepackage[english]{babel}

\usepackage[rmargin=2.5cm,
            lmargin=2.5cm,
            top=2.5cm,
            bottom=2.5cm]{geometry}
            
\usepackage{enumerate}
\usepackage{paralist}
\renewcommand{\labelenumi}{(\theenumi)}

\usepackage{graphicx}
\usepackage{subfig}
\usepackage{placeins}

\usepackage{amsmath}
\usepackage{amsfonts}
\usepackage{amssymb}
\usepackage{dsfont}
\usepackage{mathtools}

\usepackage{longtable}
\usepackage{multirow}

\usepackage[colorlinks=false, pdfborder={0 0 0}]{hyperref}

\usepackage{natbib}
\usepackage{url}

\usepackage{caption}
\usepackage{framed}

\newcommand{\vectheta}{\boldsymbol{\theta}}

\newcommand{\by}{\boldsymbol{y}}

\newcommand{\Prob}{\mathbb{P}}

\usepackage{setspace}
\begin{document}

\begin{center}

{\Large \bfseries Iterative Estimation of Mixed Exponential Random Graph Models with Nodal Random Effects}
\vspace{5 mm}

{\large Sevag Kevork$^\star$, G\"oran Kauermann$^\star$. } \\
\vspace{2 mm}
{\textit{$^\star$Institut f\"ur Statistik, Ludwig-Maximilians-Universit\"at M\"unchen, Germany.}}

\vspace{5 mm}

\today 

\vspace{5 mm}

\end{center}

\begin{abstract}
The presence of unobserved node specific heterogeneity in Exponential Random Graph Models (ERGM) is a general concern, both with respect to model validity as well as estimation instability. We therefore include node specific random effects in the ERGM that account for unobserved heterogeneity in the network. This leads to a mixed model with parametric as well as random coefficients, labelled as mixed ERGM. Estimation is carried out by iterating between approximate pseudolikelihood estimation for the random effects and maximum likelihood estimation for the remaining parameters in the model. This approach provides a stable algorithm, which allows to fit nodal heterogeneity effects even for large scale networks. We also propose model selection based on the AIC to check for node specific heterogeneity.
\end{abstract}

\textbf{Keywords:} exponential random graph models; random effects; generalized linear mixed models; network data analysis

\vspace*{6pt}

\section{Introduction}
\label{sec:Introduction}

The analysis of network data has become a challenging and emerging field in statistics in the last years. \cite{GoldenbergZheng:2010}, \cite{HunterKrivitskySchweinberger:2012} and \cite{Fienberg:2012}  provide comprehensive articles on statistical approaches, challenges and developments in network data analysis. We also refer to \cite{Kolaczyk:2009} and \cite{kolaczyk2014statistical} for a general introduction and the related routines for network data analysis in R. In this paper, we concentrate on Exponential Random Graph Models (ERGM) originally introduced in \cite{FrankStrauss:1986} and more deeply discussed e.g. in \cite{LusherKoskinenRobins:2013}. Unless node specific covariates are included in the ERGM, the probability for all possible edges across the graph is assumed to be homogeneous, which also means that any permutation of the node labels will yield the same probability. This is a questionable assumption, in particular in large networks, which also contributes towards stability problems for estimation as discussed e.g. in \cite{schweinberger2011instability} or \cite{schweinberger2017foundations}. We follow \cite{thiemichen2016bayesian} and extend the ERGM by incorporating node specific heterogeneity effects to overcome the homogeneity assumption of ERGM and capture the unobserved heterogeneity in the network.\\

Consider a network of $n$ actors for which some dyadic relationships have been recorded. These relations can be represented in the form of an $n \times n$ adjacency matrix $Y$, with elements $Y_{ij} = 1$ if an edge from $i$ to $j$ exists and $Y_{ij} = 0$ otherwise. In undirected networks we have $Y_{ij} = Y_{ji} \,\forall \,i \neq j$, which we assume throughout this paper. We consider the probability of observing a given network conditional on a set of (sufficient) network statistics is given by the ERGM
 
\begin{align}\label{eq:p1}
\Prob\bigl( Y= y | \vectheta \bigr) &= \frac{\exp \{ \vectheta^{T}\, \boldsymbol{s}(y)\}}{\kappa(\vectheta)}.
\end{align}  

Here, $\boldsymbol{s}(y)$ is the vector of network statistics and $\vectheta$ is the vector of model coefficients. Vector $\boldsymbol{s}(.)$ includes any structural characteristics of the network and we refer to \cite{Snijders-etal:2006} for a general discussion on network statistics, see also \cite{HunterHandcock:2006}. The denominator $\kappa(\vectheta)$ in (\ref{eq:p1}) represents the normalizing factor to ensure that (\ref{eq:p1}) is a legitimate probability mass function. In general, $\kappa(\vectheta)$ is numerically intractable unless for miniature networks. Estimation of $\vectheta$ in model (\ref{eq:p1}) needs therefore to be carried out simulation based. An early reference for estimation of ERGMs is \cite{snijders2002markov}. For a general discussion, we refer to \cite{HunterKrivitskySchweinberger:2012}. A numerical stable routine has been proposed in \cite{hummel2012improving} using a so-called stepping algorithm. Bayesian estimation is proposed in \cite{CaimoFriel:2011}. An important property resulting from equation (\ref{eq:p1}) is that $\boldsymbol{s}(y)$ is a vector of sufficient statistics for the network. This means, two networks, which coincide in $\boldsymbol{s}(y)$, have the same probability. In particular, this means that all possible node specific heterogeneity in the network is explained by exogenous effects, which may also be included in model (\ref{eq:p1}). This can be seen as questionable assumption. For instance, in a friendship network we may believe that the formation of friendships (edges) between individuals (nodes) is driven by many factors, observable as well as unobservable. We may suspect that there are quantities, intangible factors specific to each individual (node) that are difficult if not impossible to measure. It seems therefore, advisable to include node specific heterogeneity to capture possible unobserved heterogeneity of the nodes.\\ 

An early model that incorporates node specific heterogeneity is the so-called $p_1$ model proposed in \cite{HollandLeinhardt:1981}. The model includes parametric sender and receiver effects but no network statistics (except of reciprocity). Random nodal heterogeneity was proposed by \cite{Duijn-etal:2004} or \cite{Zijlstra-etal:2006} which led to the so-called $p_2$ model. \cite{thiemichen2016bayesian} combined the approach with ERGMs and proposed Bayesian estimation, which however is infeasible for large networks, i.e. networks with more than about 100 actors. We follow the approach of \cite{thiemichen2016bayesian} and extended the ERGM towards 

\begin{align}\label{eq:extergm}
\Prob \bigl( Y = y | \vectheta, \boldsymbol{u} \bigr) &= \frac{ \exp \bigl\{ \vectheta^{T} \boldsymbol{s}( y )+ \boldsymbol{u}^{T} \boldsymbol{t}(y) \bigr\} }{ \kappa\bigl( \vectheta, \boldsymbol{u} \bigr) },
\end{align}

where $\boldsymbol{s}(y) = (s_{0}(y), s_{1}(y), \ldots)$ is, as above, the $p$ dimensional vector of network statistics with $s_{0}(y) = \sum_{i}\sum_{j>i}y_{ij}$ as intercept and $\boldsymbol{t}(y) = \bigl( \sum_{j \neq 1}y_{1j}, \, \sum_{j \neq 2}y_{2j}, \ldots \bigr)$ is the $n$ dimensional vector of node degrees. The normalization now equals $\kappa(\vectheta, \boldsymbol{u}) = \log \sum_{\by \in \mathcal{Y}} \exp \bigl(\vectheta^{T}\boldsymbol{s}(y) + \boldsymbol{u}^{T}\boldsymbol{t}(y)  \bigr)$, where $\mathcal{Y}$ is the set of $n$ by $n$ networks. Conditional on $\boldsymbol{u}$, we obtain node specific heterogeneity, which can be seen as follows. We assume now that $\vectheta \in \mathbb{R}^{p}$ is a $p$ dimensional parameter vector while $\boldsymbol{u} = (u_{1}, u_{2}, \ldots)$ is a $n$ dimensional vector of random node specific coefficients with

\begin{align}\label{eq:randomdist}
\boldsymbol{u} \sim \mathcal{N}(0, \, \sigma_{u}^{2} \, \boldsymbol{\mathcal{I}}_{n})
\end{align}

with $\sigma_{u}^{2}$ as variance and $\boldsymbol{\mathcal I}_n$ as $n$ dimensional identity matrix. This leads to a mixed model with fixed and random coefficients, termed in the following as mixed ERGM, or in short mERGM. The reasoning behind the model structure can be seen through the conditional model for a single edge $Y_{ij}$ conditional on the rest of the network denoted as $Y_{-ij}$. From (\ref{eq:extergm}) we obtain

\begin{align}\label{eq:logextergm}
\log \Biggl\{ \frac{\Prob(Y_{ij}=1|y_{-ij}, \vectheta, \boldsymbol{u})}{\Prob(Y_{ij}=0|y_{-ij}, \vectheta, \boldsymbol{u})}\Biggr\} &= \vectheta^{T}\, \Delta_{ij} \, \boldsymbol{s}(y) \, + u_{i} + u_{j}.
\end{align}

Here $\Delta_{ij}\boldsymbol{s}(y) = \boldsymbol{s}(y_{ij} = 1, y_{-ij}) - \boldsymbol{s}(y_{ij} = 0, y_{-ij})$ is the so-called change statistics where $y_{-ij}$ is the network except of edge $y_{ij}$. The terms $u_i$ and $u_j$ are the random node specific coefficients accounting for heterogeneity not captured with $\Delta_{ij}\boldsymbol{s}(y)$. If we assume normality for coefficients $u_{i}$, formula (\ref{eq:logextergm}) resembles a mixed logistic regression model as extensively discussed e.g. in \cite{10.2307/2290687}. A similar model to (\ref{eq:logextergm}) has been proposed by  \cite{box-steffensmeier_christenson_morgan_2018} taking the coefficients $\boldsymbol{u}$ in (\ref{eq:logextergm}) as random normal variable with mean zero and unknown variance. For estimation they apply a pseudolikelihood approach. Though this circumvents the numerical burden of estimation in ERGM, it comes for the price of biased estimation  of the paramteric coefficients $\vectheta$. In other words, their estimation approach is biased even if the random node effects are close or equal to zero. We refer to \cite{schmid2017exponential} for a general discussion on pseudolikelihood estimation in ERGMs, see also \cite{StraussIkeda:1990} or \cite{DESMARAIS20121865}.\\

Model (\ref{eq:extergm}) combined with the probability model (\ref{eq:randomdist}) can also be seen as special case of the Exponential-family Random Network Model as proposed in \cite{fellows2012exponential}. They propose simulation based estimation, which is restricted to small network sizes. The restriction to small networks is also pointed out in \cite{thiemichen2016bayesian}. Hence, even though model (\ref{eq:extergm}) can be considered as ERGM with some prior structure on coefficients $\boldsymbol{u}$ given in (\ref{eq:extergm}), estimation based on simulation based methods (see \citealt{snijders2002markov}) becomes infeasible for larger networks. We therefore, propose a different estimation strategy, which is motivated in the next paragraph. At this point, we also note that several approaches to capture unobserved heterogeneity were introduced in network data analysis, for instance, \cite{koskinen2009using} introduced binary latent class indicators, \cite{schweinberger2015local} examined local dependence using a Bayesian framework in random graph models and \cite{henry2020modeling} developed a modeling framework to capture unobserved heterogeneity in the effects of nodal covariates.\\

The first goal of this paper is to provide an iterative estimation strategy, combining both maximum likelihood and pseudolikelihood estimation techniques. To be more specific, we take model (\ref{eq:logextergm}) as starting point and make use of pseudolikelihood estimation for the random coefficients $\boldsymbol{u}$ while for estimation of $\vectheta$ we use the steplength MCMC-MLE approach proposed in \cite{hummel2012improving} and the corresponding stepping algorithm \cite{hummel2012improving}, implemented in the \texttt{ergm} package in R (see \citealt{HunterHandcock-etal:2008}). These two steps are used iteratively, leading to feasible estimation. Our estimation strategy allows us to fit large scale networks, and we observe that the inclusion of the nodal effects works towards numerical estimation stability, as demonstrated through examples and simulations. Moreover, as the second goal of this paper, we propose a simple model selection strategy to evaluate nodal heterogeneity. To be specific, we use Akaike's Information Criterion (see \citealt{akaike1974new}) to select a model with or without nodal effects. The latter is calculated numerically by employing a Laplace approximation.\\

This paper is organized as follows. In Section \ref{sec:estimation} we discuss the estimation for the underlying model in detail and introduce our algorithm. Furthermore, in Section \ref{sec:selection} we present a simulation based method for evaluating our model, which allows us to calculate the AIC value for the mERGM and compare it with the AIC value of a corresponding ERGM. In Section \ref{sec:simulation}, we present a simulation based study with the corresponding results, in Section \ref{sec:examples}, we then apply our approach to three data examples. Finally, Section \ref{sec:discussion} closes with a discussion.

\section{Model and Estimation}{}
\label{sec:estimation}
We consider model (\ref{eq:extergm}) where the nodal heterogeneity effects $u_{1}, \ldots, u_{n}$ are assumed to be random with $\sigma_{u}^{2}$ as variance and $\boldsymbol{\mathcal I}_n$ as $n$ dimensional identity matrix. The aim is to fit parameter $\vectheta$ taking nodal heterogeneity into account. Moreover, we need to estimate $\sigma_{u}^{2}$, which in fact quantifies the amount of nodal heterogeneity.\\ 

In principle, we need to maximize the marginal log-likelihood

\begin{align}\label{eq:marginalloglik}
l(\vectheta, \, \sigma^{2}_{u}) &= \log {\displaystyle \int \frac{ \exp \bigl\{ \vectheta^{T} \boldsymbol{s}(y)+ \boldsymbol{u}^{T} \boldsymbol{t}(y) \bigr\} }{ \kappa\bigl( \vectheta, \boldsymbol{u} \bigr) }} \cdot \frac{1}{(2 \pi \sigma_{u}^{2})^{\frac{n}{2}}} \cdot\exp\Bigl( -\frac{1}{2} \, \frac{\boldsymbol{u}^{T}\boldsymbol{u}}{\sigma^{2}_{u}} \Bigr) \prod_{i=1}^{n} \, du_{i} \notag\\
&= \vectheta^{T} \boldsymbol{s}(y) - \frac{n}{2}\log(2\pi) - \frac{n}{2}\log(\sigma_{u}^{2}) + \log {\displaystyle \int \exp \Bigl( g(\boldsymbol{u}, \vectheta, \sigma^{2}_{u}) \Bigr)} \prod_{i=1}^{n} du_{i}
\end{align}

where

\begin{align}\label{eq:gfunction}
g(\boldsymbol{u}, \vectheta, \sigma^{2}_{u}) = \boldsymbol{u}^{T} \boldsymbol{t}(y) - \log(\kappa(\vectheta, \boldsymbol{u})) - \frac{1}{2} \cdot \frac{\boldsymbol{u}^{T}\boldsymbol{u}}{\sigma^{2}_{u}}.
\end{align}

We may approximate the integral in (\ref{eq:marginalloglik}) using a Laplace approximation. Let therefore $\boldsymbol{\hat{u}}$ be the maximizer of $g(\boldsymbol{u}, \vectheta, \sigma^{2}_{u})$, which apparently depends on both, $\vectheta$ and $\sigma^{2}_{u}$. This leads to the approximate log-likelihood 

\begin{align}\label{eq:laplace}
l(\vectheta, \sigma^{2}_{u}) \approx \vectheta^{T} \boldsymbol{s}(y) + \boldsymbol{\hat{u}}^{T} \boldsymbol{t}(y) - \log(\kappa(\vectheta, \boldsymbol{\hat{u}})) - \frac{1}{2} \cdot \frac{\boldsymbol{\hat{u}}^{T}\boldsymbol{\hat{u}}}{\sigma^{2}_{u}} - \frac{n}{2}\log(2\pi) - \frac{n}{2}\log(\sigma_{u}^{2}) - \frac{1}{2} \, \log \left| \frac{\partial^{2} g(\boldsymbol{\hat{u}}, \vectheta, \sigma^{2}_{u})}{\partial u \, \partial u} \right|.
\end{align}

Note that the likelihood can be considered as profile likelihood, where $\boldsymbol{u}$ is "estimated" through maximizing (\ref{eq:gfunction}). If we now treat $\boldsymbol{\hat{u}}$ as given, then maximization of $l(\vectheta, \sigma_{u}^{2})$ with respect to $\vectheta$ corresponds to maximizing the likelihood of the probability model

\[
\Prob \bigl( Y = y | \vectheta, \boldsymbol{\hat{u}} \bigr) = \frac{ \exp \bigl\{ \vectheta^{T} \boldsymbol{s}(y)+ \boldsymbol{\hat{u}}^{T} \boldsymbol{t}(y) \bigr\} }{ \kappa\bigl( \vectheta, \boldsymbol{\hat{u}} \bigr) },
\]

where $\boldsymbol{\hat{u}}^{T}\boldsymbol{t}(y)$ is fixed as given offset. The terminology offset means here that the quantity $\boldsymbol{\hat{u}}^{T}\boldsymbol{t}(y)$ is treated as fixed. In other words, setting the random coefficients to $\boldsymbol{\hat{u}}$ simplifies the estimation of $\vectheta$ to Maximum Likelihood estimation in an ERGM with offset $\boldsymbol{\hat{u}}^{T}\boldsymbol{t}(y)$. This is numerically available with standard software packages (e.g. \citealt{HunterHandcock-etal:2008}) and the stepping algorithm proposed in \cite{hummel2012improving}. Let us therefore first look in more detail how to obtain $\boldsymbol{\hat{u}}$ if we keep $\vectheta$ as fixed. Note that $\boldsymbol{\hat{u}}$ results by solving

\[
\frac{\partial \, g(\boldsymbol{u}, \vectheta, \sigma^{2}_{u}) }{\partial \boldsymbol{u}} = 0.
\]

Apparently, this is numerically problematic, since $\kappa(\vectheta, \boldsymbol{u})$ is numerically intractable. Differentiation yields

\[ \frac{\partial \log \kappa(\vectheta, \boldsymbol{u})}{\partial \boldsymbol{u}} = \mathbb{E}(\boldsymbol{t}(y) | \boldsymbol{u})\]

which in principle can be approximated using simulation based approaches (see \citealt{snijders2002markov}). However, this is a numerically challenging task, since $\boldsymbol{u}$ is high dimensional, namely $n$ dimensional. We therefore propose to approximate the estimation step of $\boldsymbol{u}$ by pseudolikelihood estimation. To do so we look at the model for a single edge given the rest of the network. That is to say we take model (\ref{eq:logextergm}), but now we fix $o_{ij} \coloneqq \vectheta^{T}\, \Delta_{ij} \, \boldsymbol{s}(y)$ as offset and ignore the dependence on $\by$. This leads to a pseudolikelihood approach as discussed for instance in \cite{VANDUIJN200952}. The pseudo-log-likelihood thereby equals 

\begin{align}\label{eq:pseudo_loglik1}
l_{\text{pseudo}}(\boldsymbol{u}) = \sum_{i}\sum_{j>i} y_{ij} \{ o_{ij} + u_{i} + u_{j}\} - \log\{ 1 + \exp(o_{ij} + u_{i} + u_{j}) \}
\end{align}

where a priori coefficient $\boldsymbol{u}$ follows a normal distribution as given in (\ref{eq:randomdist}). The pseudo-log-likelihood (\ref{eq:pseudo_loglik1}) together with (\ref{eq:randomdist}) leads to (pseudo) Generalized Linear Mixed Effects Model, so that we can borrow estimation strategies from this field. In particular we make use of \cite{10.2307/2290687} who propose to approximate the resulting marginal likelihood using a Laplace approximation, similar to (\ref{eq:laplace}) above. This allows to estimate the a priori variance $\sigma^{2}_{u}$ and predict the random coefficients $\boldsymbol{u}$. Note, if $o_{ij}$ is independent of $y_{-ij}$, then the Laplace approach (\ref{eq:laplace}) is equal to the estimation proposed in \cite{10.2307/2290687}. Apparently, this is the case for the so-called $p_2$ model (see \citealt{Duijn-etal:2004}). To be specific, the Laplace approximated pseudolikelihood with $o_{ij} = (o_{1\,2}, o_{1\,3}, \ldots , o_{n-1\,n})$ as offset and constant terms omitted results through 

\begin{align}\label{eq:pseudo_loglik2}
l_{\text{pseudo}}(\sigma^{2}_{u}) \approx -\frac{1}{2} \cdot \frac{\boldsymbol{\check{u}}^T \boldsymbol{\check{u}}}{\sigma^{2}_{u}} - \frac{n}{2} \log(\sigma^{2}_{u})
\end{align}

where $\boldsymbol{\check{u}}$ is the maximizer of 

\[ 
\check{g}(\boldsymbol{u}, \sigma^{2}_{u} ;o) = \boldsymbol{u}^{T}\boldsymbol{t}(y) - \sum_{i}\sum_{j} \log \{ 1 + \exp(o_{ij} + u_{i} + u_{j}) \} - \frac{1}{2} \cdot \frac{\boldsymbol{u}^{T}\boldsymbol{u}}{\sigma^{2}_{u}}
\]

We call $\boldsymbol{\check{u}}$ an estimate subsequently, even though of course it is a predictor given that $\boldsymbol{u}$ is considered as random. This is implemented in multiple R packages, see e.g. \cite{faraway2016extending}. To be specific, for estimation of $\boldsymbol{u}$ we use the {\tt mgcv} package (see \citealt{mgcv}). Denoting with $\check{\boldsymbol{u}}=(\check{u}_{1}, \ldots , \check{u}_{n})$ the resulting estimates, we set $\check{\boldsymbol{u}}^{T} \boldsymbol{t}(y)$ in (\ref{eq:extergm}) as offset and estimate parameter $\vectheta$ using simulation based techniques. For this step we use the \texttt{ergm} package (see \citealt{HunterHandcock-etal:2008}). Both estimation steps are used iteratively until convergence. That is we take the current estimate $\hat{\vectheta}_{(t)}$ and update $\check{\boldsymbol{u}}$ with pseudolikelihood leading to $\check{\boldsymbol{u}}_{(t+1)}$. This in turn allows to update $\hat{\vectheta}$ after replacing the offset by $\check{\boldsymbol{u}}_{(t+1)}^{T} \boldsymbol{t}(y)$. Our algorithmic steps work in detail as follows:

\begin{framed}
\textbf{Algorithm:} Fit ERGM with nodal random effect components
\hrule
\begin{enumerate}
	\setcounter{enumi}{-1}
\renewcommand{\labelenumi}{\textit{Step \arabic{enumi}:}}
\item Obtain a prediction for $\boldsymbol{u}$ and estimate $\sigma^{2}_{u}$:
	\begin{enumerate}[(i)]
		\item Fit the model \texttt{logit} $\mathbb{P}(y_{ij} = 1|y_{-ij}, \vectheta, \boldsymbol{u}) = \theta_{(0)} + u_{i} + u_{j}$ to the data, where $1 \le i < j \le n$
		\item extract the vector of the predicted random effects $\boldsymbol{\check{u}}_{(0)}$ as offset and set $t=0$
	\end{enumerate}
\item Estimate $\vectheta$ with ERGM and take $\boldsymbol{\check{u}}_{(t)}^{T} \boldsymbol{t}(y)$ as an offset parameter:
	\begin{enumerate}[(i)]
		\item Fit the model $\mathbb{P}(Y=y|\vectheta) \propto \text{exp}\{ \boldsymbol{\theta}_{(t+1)}^{T} \boldsymbol{s}(y) + \underbrace{\boldsymbol{\check{u}}_{(t)}^{T} \boldsymbol{t}(y)}_{\text{offset}} \}$ using maximum likelihood and simulation based methods
		\item extract $o_{ij}=\boldsymbol{\theta}_{(t+1)}^{T}\Delta_{ij}\boldsymbol{s}(y)$ as new offset for $1 \le i < j \le n$
	\end{enumerate}
\item Update $\boldsymbol{\check{u}}_{(t+1)}$ and $\hat{\sigma}^{2}_{u(t+1)}$ now taking $o_{ij}$ as offset parameter:
	\begin{enumerate}[(i)]
		\item Fit the model \texttt{logit} $\mathbb{P}(y_{ij} = 1|y_{-ij}, \vectheta, \boldsymbol{u}) = o_{ij} + u_{i} + u_{j}$ with priori structure (\ref{eq:randomdist}) to the data
		\item extract the vector of the predicted random effects $\boldsymbol{\check{u}}_{(t+1)}$ as new offset 
	\end{enumerate}
\end{enumerate}
Set $t=t+1$ and iterate between step $1$ and $2$ until $\vectheta_{(t)}$ converges. Convergence is achieved if $|\vectheta_{(t)} - \vectheta_{(t+1)}| \leq \epsilon = 0.05$.
\end{framed}

We need to mention that pseudolikelihood estimation in network data analysis is biased, in particular, if dyadic statistics are involved. Our setting here, however, is slightly different since we use pseudolikelihood estimation only for the degree statistics $\boldsymbol{t}(y)$. Moreover, we consider the coefficients $\boldsymbol{u}$ not to be fixed but random so that in general, we are more interested in the variance of coefficients $u_{i}$ and less interested in their true value. We refer to \cite{schmid2017exponential} or \cite{cranmer_desmarais_2011} for further discussion on pseudolikelihood estimation in network data analysis.

\section{Inference through Model Selection and Variance Estimation}
\label{sec:selection}

The central question in network data analysis is to explain the dominating factors in the network, i.e. the sufficient statistics describing the network structure. If we allow for node specific heterogeneity, we are additionally faced with the problem of model selection. In other words, we need to describe whether the network data at hand shows evidence with respect to heterogeneous nodes or whether the homogeneity assumption of ERGM seems valid. We tackle this question by approximate calculation of the Akaike Information Criterion (AIC). To do so, we assume for simplicity that the determinant component in (\ref{eq:laplace}) depends only weakly on $\vectheta$ so that we can ignore it subsequently. This is in line with the arguments proposed in \cite{10.2307/2290687} who suggest the use of Laplace approximation in generalized linear mixed models. Note that

\begin{align}\label{eq:determinant}
\frac{\partial^{2} \, g(\boldsymbol{u}, \vectheta, \sigma^{2}_{u}) }{\partial \boldsymbol{u} \, \partial \boldsymbol{u^T}} = - Var(\boldsymbol{t}(Y) \, | \vectheta, \boldsymbol{u}) - \frac{1}{\hat{\sigma}^{2}_{u}}\, \boldsymbol{\mathcal I}_{n}.
\end{align}

Hence, ignoring the determinant component in (\ref{eq:laplace}) is justified if we assume that the variance matrix of the degree vector $\boldsymbol{t}(Y)$ depends only weakly on $\vectheta$. We refer to \cite{10.2307/2290687} for a deeper discussion and motivation which justifies to pursue this simplification. Generally, the variance is difficult to calculate or even infeasible for large networks, so that we make use of simulations to estimate (\ref{eq:determinant}). To do so, we simulate networks in order to obtain a simulation based approximation for $\kappa(\hat{\vectheta}, \boldsymbol{\check{u}})$. We make further use of the simulated networks to obtain a simulation based approximation of $Var(\boldsymbol{t}(y)|\hat{\vectheta}, \boldsymbol{\check{u}})$. To be specific, let $y^{\star(1)}, \ldots, y^{\star(N)}$ be a set of (independent) network simulations derived from model (\ref{eq:extergm}) with $\vectheta$ set to $\hat{\vectheta}$ and $\boldsymbol{u}$ set to $\boldsymbol{\check{u}}$. We estimate $Var(\boldsymbol{t}(y) \, | \, \hat{\vectheta}, \, \check{\boldsymbol{u}})$ through

\[
\frac{1}{N} \sum_{j=1}^{N} \left[ \boldsymbol{t}(y^{\star(j)}) - \bar{t}(y^{\star}) \right] \, \left[ \boldsymbol{t}(y^{\star(j)}) - \bar{t}(y^{\star}) \right]^{T} 
\]  

where $\bar{t}(y^{\star})$ is the arithmetic mean of the simulated values.\\

With these prerequisites, we can now approximate all quantities in (\ref{eq:laplace}). This also holds for the normalization constant, which is estimated through

\[
\hat{\kappa}(\hat{\vectheta}, \boldsymbol{\check{u}}) = \frac{1}{N} \sum_{j=1}^{N} \exp\bigl( \hat{\vectheta}^{T}\boldsymbol{s}(y^{\star(j)}) \, + \, \boldsymbol{\check{u}}^{T}\boldsymbol{t}(y^{\star(j)}) \bigr).
\]

Model comparison can now be carried out with the AIC. Setting $p$ as the number of parameters in $\vectheta$ the AIC results to 

\begin{align}\label{eq:aiceq}
AIC_{mERGM} = -2 \, l(\hat{\vectheta}, \hat{\sigma}^{2}_{u}) \, + \, 2 \cdot(p + 1)
\end{align}

Note that (\ref{eq:aiceq}) resembles the marginal AIC, that is after integrating out $\boldsymbol{u}$. We refer to \cite{10.1093/biomet/asq042} or \cite{doi:10.1093/biomet/92.2.351} for a deeper discussion of applying AIC in random effects models. In our case, formula (\ref{eq:aiceq}) serves as approximation, relying on the pseudolikelihood estimation for $\boldsymbol{u}$.\\

We compare the AIC in the mERGM to the resulting AIC in the case of node homogeneity, that is by setting $\sigma_{u}^{2}=0$. This is carried out in a similar way, but we set $\boldsymbol{u} = 0$. \\
In other words we use the likelihood in (\ref{eq:p1}) by calculating $\kappa(\vectheta)$ simulation based from $N$ draws $y^{\star(1)}, \ldots, y^{\star(N)}$ from model (\ref{eq:p1}) with $\vectheta$ set to the ML estimate in model (\ref{eq:p1}). We call this

\begin{align}\label{eq:aicergm}
AIC_{ERGM} = -2 \, l_{ERGM}(\hat{\vectheta}) \, + \, 2 p
\end{align}

where $l_{ERGM}$ is the log-likelihood in the ERGM resulting when $\boldsymbol{u} \equiv 0$.\\

Though the focus of the paper lies on model comparison, we shortly discuss how to calculate the variance of the estimate if the algorithm above is used. In the ERGM, we obtain the Fisher information matrix

\begin{align}\label{eq:varergm}
I(\vectheta) = Var(\boldsymbol{s}(Y)|\vectheta).
\end{align}

This can be estimated simulation based, that is we simulate from model (\ref{eq:p1}) and calculate $Var(\boldsymbol{s}(Y))$ based on the simulated values. For the mERGM we need to take into account that coefficients $\boldsymbol{u}$ are considered to be random so that in principle, we need to calculate the (inverse) Fisher information of the log-likelihood. Assuming that the determinant in (\ref{eq:laplace}) does depend only weekly on $\vectheta$ and ignoring for simplicity the dependence of $\boldsymbol{\check{u}}$ on $\vectheta$ we obtain again (\ref{eq:varergm}). This is, of course, an approximation since we ignored the dependence of $\boldsymbol{\check{u}}$ and $\vectheta$ as well as estimation variability of $\sigma_{u}^{2}$. In other words, exact variance calculation in the mERGM is complicated and we here provide a rough approximation only. However, inference can be carried out by model selection via the AIC, which is what we pursue in simulations and data examples below. Still, we can make use of the simulations from above to obtain an estimate of the Fisher information and hence a variance estimate for the estimates.

\section{Simulation Study}
\label{sec:simulation}

In our simulation study we want to explore the estimation results of the model parameters and the model selection step based on two network sizes. Small networks with 50 nodes, and large networks with 500 nodes. For each network size we use network settings with different levels of nodal heterogeneity $\sigma^{2}_{u}$. For each network setting and network size we simulate 50 networks using the simulation routines from the \texttt{ergm} package \citep{HunterHandcock-etal:2008}. Each network setting has the same structural effects $\vectheta$, where $\vectheta = (\theta_{edges}, \, \theta_{gwesp}, \, \theta_{2-stars}) = (-1, \, 0.2, \, -0.3)$, but the nodal heterogeneity takes six different levels $\sigma^{2}_{u} = (0, \, 0.1, \, 0.2, \, 0.5, \, 0.8, \, 1)$, where $\boldsymbol{u}$ is randomly drawn for each simulated network and setting from a normal distribution following (\ref{eq:randomdist}). For each network size of these six heterogeneity levels we fit an ERGM and a mERGM to the 50 simulated networks. Note that the tuning parameters for the $\theta_{gwesp}$ term are the same for both ERGMs and mERGMs with (\texttt{decay = 0.8, fixed = TRUE}). Additionally, we provide the results of two more simulation studies as supplementary material, where we perturb the parameter vector $\vectheta$ with different configurations for both small and large networks. Furthermore, we also provide supplementary material illustrations of how well $\sigma^{2}_{u}$ is recovered in the simulation study. 

Our first focus is on the performance of the estimates. In Table \ref{tab:summary_sim_small} and \ref{tab:summary_sim_large} we summarize the results of our simulation study for small (50 nodes) and large (500 nodes) networks respectively, distinguishing all six different levels of nodal heterogeneity. Let us first look at the case $\sigma_{u}^{2}=0$, the fitted parametric coefficients of the ERGMs in both cases, small and large networks, show a stable estimation performance; however, this stability is more evident in large networks. The mERGMs, on the other hand, show some estimation variability and are outperformed by ERGMs concerning parameter estimation. The trend changes increasingly in favour of mERGMs, with increasing heterogeneity. At a heterogeneity of $\sigma_{u}^{2}=0.8$, the mERGMs excel with better results in small and large networks than the ERGMs, which show substantial variability in the results. At a heterogeneity of $\sigma_{u}^{2}=1$, the results indicate severe stability problems of the ERGM estimates, especially in the small networks, but the more the size of the network increases, the less unstable the estimates of the ERGMs become. This fact is not surprising since a heterogeneity of $\sigma_{u}^{2}=1$ in a network with $50$ nodes is a different claim than in a network with $500$ nodes. Nevertheless, the mERGMs in both cases, small and large networks, show an appropriate and stable performance. Hence, including nodal heterogeneity in the model increases the stability of the mERGM fit. This is a welcome effect of the model extension from ERGM to mERGM. 

As a second point, we consider the performance of the model selection based on the Akaike information criterion. To do so, we calculate for each of the $50$ simulations in each setting the log ratio 

 \[ \log \Biggl( \frac{AIC_{mERGM_{k}}}{AIC_{ERGM_{k}}} \Biggr)  \;\;\; k = 1, \ldots, 50. \]

\begin{figure}[!htbp]\centering
  \includegraphics[width=\textwidth]{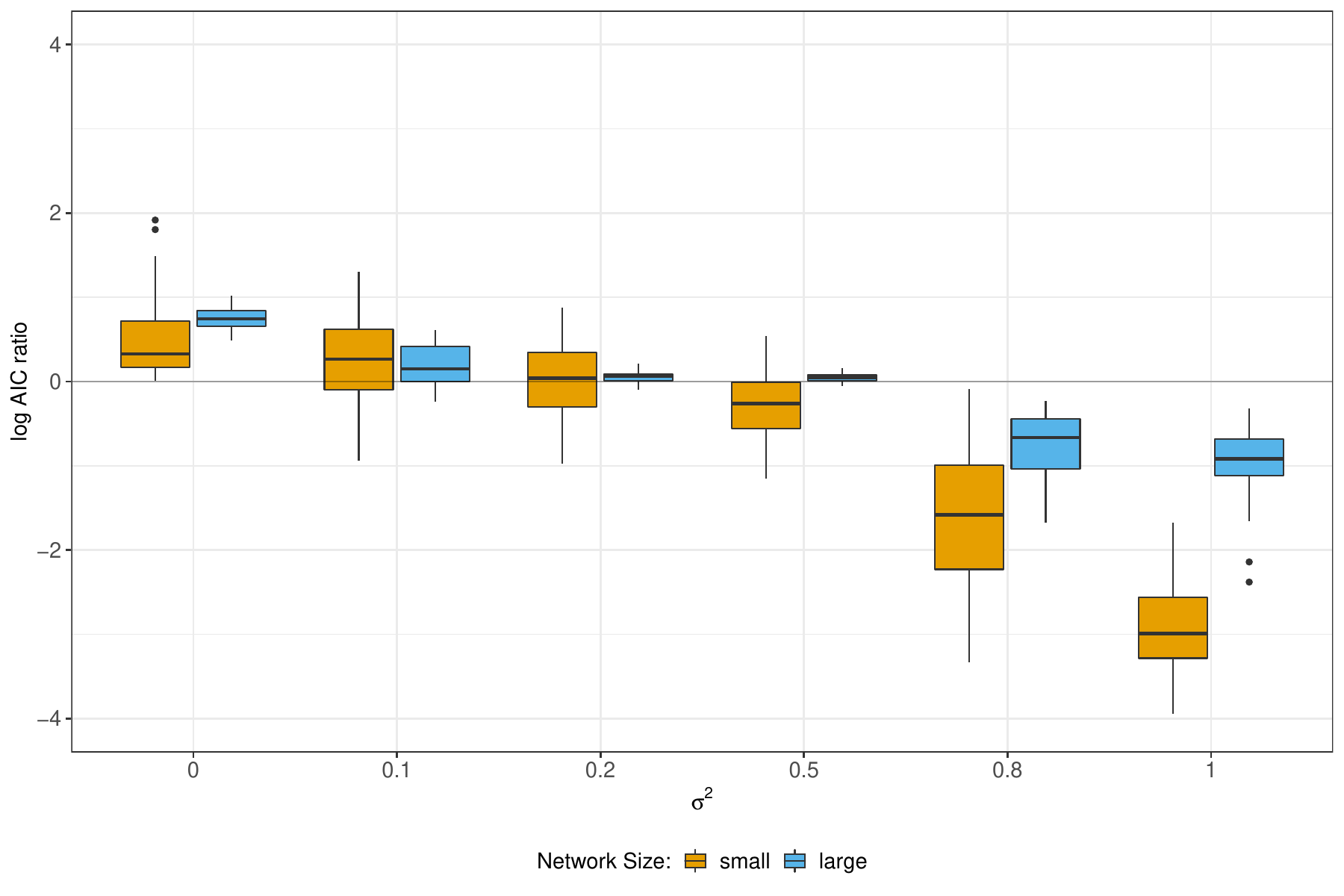}
\caption{Resulting log AIC Ratios of mERGM and ERGM for each network setting and size. Orange boxplots indicate small networks (50 nodes), blue boxplots large networks (500 nodes).}
\label{fig:ergm_mergm_aic}
\end{figure}

 If the log ratio is positive, it speaks in favour of a model without nodal heterogeneity. In contrast, if the log ratio is negative, there is an indication of model heterogeneity. Figure \ref{fig:ergm_mergm_aic} visualizes the log AIC ratio for different strengths of nodal heterogeneity and both network sizes. In small (50 nodes) and large (500 nodes) networks, the ERGM was correctly preferred in the case of missing nodal heterogeneity, that is $\sigma_{u}^{2}=0$. With increasing nodal heterogeneity level in the network, the mERGM becomes more appropriate, and from a heterogeneity level of $\sigma_{u}^{2}=0.5$ and $\sigma_{u}^{2} = 0.8$ for small networks and large networks, respectively, the mERGM gets clearly selected based on the AIC. We, therefore, can conclude that the AIC allows for model selection in case of node specific heterogeneity. 

\begin{table}[htb!]\centering
\small
\begin{tabular}{cllcccccc}
\hline\\[-1ex]
\multicolumn{9}{c}{Network Size: 50 Nodes}\\[1ex]
\hline\\[-1ex]
$\sigma_{u}^2$ & Model type & Parameter & Real Value & Mean & SD & Q 0.1 & Median & Q 0.9 \\[1ex]
\hline\hline
\multirow{3}{*}{0} &\multirow{3}{*}{ERGM} & $\theta_{edges}$ & -1 & -0.71 & 0.79 & -1.65 & -0.74 & 0.37 \\
& & $\theta_{gwesp}$ & 0.2 & 0.06 & 0.78 & -0.06 & 0.17 & 0.38 \\
& & $\theta_{2-stars}$ & -0.3 & -0.35 & 0.13 & -0.50 & -0.34 & -0.20  \\
\hline
\multirow{3}{*}{0} &\multirow{3}{*}{mERGM} & $\theta_{edges}$ & -1 & 0.52 & 1.18 & -0.89 & 0.39 & 0.31 \\
& & $\theta_{gwesp}$ & 0.2 & 0.06 & 0.81 & -0.06 & 0.17 & 0.38 \\
& & $\theta_{2-stars}$ & -0.3 & -0.58 & 0.19 & -0.86 & -0.57 & -0.35 \\
\hline\hline
\multirow{3}{*}{0.1} &\multirow{3}{*}{ERGM} & $\theta_{edges}$ & -1 & -1.11 & 0.77 & -2.13 & -1.12 & -0.15 \\
& & $\theta_{gwesp}$ & 0.2 & 0.16 & 0.16 & 0.03 & 0.19 & 0.34 \\
& & $\theta_{2-stars}$ & -0.3 & -0.25 & 0.12 & -0.39 & -0.24 & -0.10 \\
\hline
\multirow{3}{*}{0.1} &\multirow{3}{*}{mERGM} & $\theta_{edges}$ & -1 & 0.06 & 0.89 & -0.96 & -0.03 & 0.98 \\
& & $\theta_{gwesp}$ & 0.2 & 0.16 & 0.16 & -0.02 & 0.19 & 0.34 \\
& & $\theta_{2-stars}$ & -0.3 & -0.45 & 0.14 & -0.60 & -0.44 & -0.28 \\
\hline\hline
\multirow{3}{*}{0.2} &\multirow{3}{*}{ERGM} & $\theta_{edges}$ & -1 & -1.63 & 0.52 & -2.21 & -1.72 & -0.88 \\
& & $\theta_{gwesp}$ & 0.2 & 0.17 & 0.21 & -0.06 & 0.17 & 0.39 \\
& & $\theta_{2-stars}$ & -0.3 & -0.17 & 0.08 & -0.31 & -0.16 & -0.06 \\
\hline
\multirow{3}{*}{0.2} &\multirow{3}{*}{mERGM} & $\theta_{edges}$ & -1 & -0.73 & 0.61 & -1.44 & -0.85 & 0.21 \\
& & $\theta_{gwesp}$ & 0.2 & 0.18 & 0.21 & -0.07 & 0.19 & 0.40 \\
& & $\theta_{2-stars}$ & -0.3 & -0.33 & 0.11 & -0.49 & -0.31 & -0.20 \\
\hline\hline
\multirow{3}{*}{0.5} &\multirow{3}{*}{ERGM} & $\theta_{edges}$ & -1 & -2.05 & 0.43 & -2.51 & -2.10 & -1.47 \\
& & $\theta_{gwesp}$ & 0.2 & 0.21 & 0.12 & 0.03 & 0.21 & 0.37 \\
& & $\theta_{2-stars}$ & -0.3 & -0.07 & 0.06 & -0.13 & -0.06 & -0.01 \\
\hline
\multirow{3}{*}{0.5} &\multirow{3}{*}{mERGM} & $\theta_{edges}$ & -1 & -0.71 & 0.62 & -1.35 & -0.77 & 0.03 \\
& & $\theta_{gwesp}$ & 0.2 & 0.21 & 0.12 & 0.03 & 0.22 & 0.36 \\
& & $\theta_{2-stars}$ & -0.3 & -0.26 & 0.08 & -0.38 & -0.26 & -0.17 \\
\hline\hline
\multirow{3}{*}{0.8} &\multirow{3}{*}{ERGM} & $\theta_{edges}$ & -1 & -2.90 & 0.27 & -3.17 & -2.93 & -2.58 \\
& & $\theta_{gwesp}$ & 0.2 & 0.21 & 0.14 & -0.01 & 0.21 & 0.34 \\
& & $\theta_{2-stars}$ & -0.3 & 0.02 & 0.04 & -0.04 & 0.03 & 0.07 \\
\hline
\multirow{3}{*}{0.8} & \multirow{3}{*}{mERGM} & $\theta_{edges}$ & -1 & -1.02 & 0.22 & -1.33 & -1.04 & -0.74 \\
& & $\theta_{gwesp}$ & 0.2 & 0.20 & 0.16 & -0.03 & 0.23 & 0.37 \\
& & $\theta_{2-stars}$ & -0.3 & -0.39 & 0.14 & -0.49  & -0.33 & -0.29 \\
\hline\hline
\multirow{3}{*}{1} &\multirow{3}{*}{ERGM} & $\theta_{edges}$ & -1 & 1123.83 & 9739.40 & -186.45 & -12.76 & 4.27 \\
& & $\theta_{gwesp}$ & 0.2 & -1112.36 & 4651.95 & -737.19 & -15.23 & 0.36 \\
& & $\theta_{2-stars}$ & -0.3 & 785.38 & 10336.06 & -2118.73 & -3.77 & 104.20  \\
\hline
\multirow{3}{*}{1} & \multirow{3}{*}{mERGM} & $\theta_{edges}$ & -1 & -1.23 & 0.40 & -1.68 & -1.23 & -0.74 \\
& & $\theta_{gwesp}$ & 0.2 & 0.26 & 0.18 & 0.01 & 0.30 & 0.49 \\
& & $\theta_{2-stars}$ & -0.3 & -0.37 & 0.10 & -0.49 & -0.35 & -0.26 \\
\hline\\[-1ex]
\end{tabular}
\caption{\label{tab:summary_sim_small} Resulting means, standard deviations, the medians, 0.1 and 0.9 quantiles of the estimated coefficients of network size 50 nodes and for all six $\sigma_{u}^2$ levels.}
\end{table}

\begin{table}[htb!]\centering
\small
\begin{tabular}{cllcccccc}
\hline\\[-1ex]
\multicolumn{9}{c}{Network Size: 500 Nodes}\\[1ex]
\hline\\[-1ex]
$\sigma_{u}^2$ & Model type & Parameter & Real Value & Mean & SD & Q 0.1 & Median & Q 0.9 \\[1ex]
\hline\hline
\multirow{3}{*}{0} &\multirow{3}{*}{ERGM} & $\theta_{edges}$ & -1 & -0.96 & 0.29 & -1.31 & -0.98 & -0.55 \\
& & $\theta_{gwesp}$ & 0.2 & 0.19 & 0.05 & 0.12 & 0.20 & 0.27  \\
& & $\theta_{2-stars}$ & -0.3 & -0.30 & 0.03 & -0.34 & -0.30 & -0.27  \\
\hline
\multirow{3}{*}{0} &\multirow{3}{*}{mERGM} & $\theta_{edges}$ & -1 & 0.18 & 0.51 & -0.53 & 0.27 & 0.81 \\
& & $\theta_{gwesp}$ & 0.2 & 0.19  & 0.05 & 0.13 & 0.20 & 0.27 \\
& & $\theta_{2-stars}$ & -0.3 & -0.16 & 0.04 & -0.21 & -0.16 & -0.11 \\
\hline\hline
\multirow{3}{*}{0.1} &\multirow{3}{*}{ERGM} & $\theta_{edges}$ & -1 & -1.99 & 0.24 & -2.34 & -1.99 & -1.73 \\
& & $\theta_{gwesp}$ & 0.2 & 0.19 & 0.05 & 0.13 & 0.20 & 0.25 \\
& & $\theta_{2-stars}$ & -0.3 & -0.22 & 0.02 & -0.23 & -0.21 & -0.18 \\
\hline
\multirow{3}{*}{0.1} &\multirow{3}{*}{mERGM} & $\theta_{edges}$ & -1 & -0.13 & 0.07 & -0.22 & -0.14 & -0.04 \\
& & $\theta_{gwesp}$ & 0.2 & 0.19 & 0.05 & 0.14 & 0.20 & 0.25 \\
& & $\theta_{2-stars}$ & -0.3 & -0.16 & 0.09 & --0.28 & -0.17 & -0.03 \\
\hline\hline
\multirow{3}{*}{0.2} &\multirow{3}{*}{ERGM} & $\theta_{edges}$ & -1 & -2.59 & 0.29 & -2.93 & -2.63 & -2.16 \\
& & $\theta_{gwesp}$ & 0.2 & 0.19 & 0.06 & 0.12 & 0.20 & 0.26 \\
& & $\theta_{2-stars}$ & -0.3 & -0.16 & 0.02 & -0.19 & -0.15 & -0.13 \\
\hline
\multirow{3}{*}{0.2} &\multirow{3}{*}{mERGM} & $\theta_{edges}$ & -1 & -0.57 & 0.26 & -0.92 & -0.54 & -0.22 \\
& & $\theta_{gwesp}$ & 0.2 & 0.19 & 0.07 & 0.12 & 0.19 & 0.27 \\
& & $\theta_{2-stars}$ & -0.3 & -0.45 & 0.04 & -0.49 & -0.45 & -0.42 \\
\hline\hline
\multirow{3}{*}{0.5} &\multirow{3}{*}{ERGM} & $\theta_{edges}$ & -1 & -3.65 & 0.16 & -3.82 & -3.68 & -3.45 \\
& & $\theta_{gwesp}$ & 0.2 & 0.19 & 0.04 & 0.14 & 0.20 & 0.25 \\
& & $\theta_{2-stars}$ & -0.3 & -0.06 & 0.01 & -0.08 & -0.06 & -0.05 \\
\hline
\multirow{3}{*}{0.5} &\multirow{3}{*}{mERGM} & $\theta_{edges}$ & -1 & -0.48 & 0.28 & -0.79 & -0.48 & -0.19 \\
& & $\theta_{gwesp}$ & 0.2 & 0.19 & 0.05 & 0.14 & 0.18 & 0.25 \\
& & $\theta_{2-stars}$ & -0.3 & -0.34 & 0.02 & -0.37 & -0.35 & -0.32 \\
\hline\hline
\multirow{3}{*}{0.8} & \multirow{3}{*}{ERGM} & $\theta_{edges}$ & -1 & -4.29 & 0.11 & -4.41 & -4.29 & -4.15 \\
& & $\theta_{gwesp}$ & 0.2 & 0.19 & 0.04 & 0.14 & 0.20 & 0.24 \\
& & $\theta_{2-stars}$ & -0.3 & -0.01 & 0.01 & -0.02 & -0.01 & -0.001  \\
\hline
\multirow{3}{*}{0.8} & \multirow{3}{*}{mERGM} & $\theta_{edges}$ & -1 & -1.45 & 0.24 & -1.77 & -1.45 & -1.16 \\
& & $\theta_{gwesp}$ & 0.2 & 0.20 & 0.04 & 0.15 & 0.20 & 0.24 \\
& & $\theta_{2-stars}$ & -0.3 & -0.26 & 0.02 & -0.28 & -0.26 & -0.23 \\
\hline\hline
\multirow{3}{*}{1} & \multirow{3}{*}{ERGM} & $\theta_{edges}$ & -1 & -4.58 & 0.08 & -4.67 & -4.57 & -4.48 \\
& & $\theta_{gwesp}$ & 0.2 & 0.19 & 0.04 & 0.15 & 0.19 & 0.23 \\
& & $\theta_{2-stars}$ & -0.3 & 0.01 & 0.01 & 0.01 & 0.01 & 0.02  \\
\hline
\multirow{3}{*}{1} & \multirow{3}{*}{mERGM} & $\theta_{edges}$ & -1 & -1.22 & 0.21 & -1.84 & -1.41 & -1.29 \\
& & $\theta_{gwesp}$ & 0.2 & 0.19 & 0.05 & 0.14 & 0.19 & 0.24 \\
& & $\theta_{2-stars}$ & -0.3 & -0.26 & 0.02 & -0.26 & -0.25 & -0.23 \\
\hline\\[-1ex]
\end{tabular}
\caption{\label{tab:summary_sim_large} Resulting means, standard deviations, the medians, 0.1 and 0.9 quantiles of the estimated coefficients of network size 500 nodes and for all six $\sigma_{u}^2$ levels.}
\end{table}

Additionally, we compare our approach with the approach proposed by \cite{box-steffensmeier_christenson_morgan_2018} (FERGM) to illustrate the point of biased estimation of the parametric coefficients $\vectheta$ we mentioned in Section (\ref{sec:Introduction}). We apply the same simulation approach as above, distinguishing the size of the networks and compare the results of the mERGM with them of FERGM fitted to the 50 simulated networks with setting $\sigma_{u}^{2}=1$. The results of our comparison are given in Table \ref{tab:summary_table_mergm_fergm}. As we can conclude from Table (\ref{tab:summary_table_mergm_fergm}), regardless of the network size, the performance of FERGM is worse. Nevertheless, we have to note that \cite{box-steffensmeier_christenson_morgan_2018} also points out that their approach might only be suitable for certain research questions or certain network types.  

\begin{table}[htb!]\centering
\footnotesize 
\begin{tabular}{lllcccccc}
\hline\\[-1ex]
Network Size & Model type & Parameter & Real Value & Mean & SD & Q 0.1 & Median & Q 0.9  \\[1ex]
\hline\hline\\[-1ex]
\multirow{3}{*}{50 Nodes} & \multirow{3}{*}{mERGM} & $\theta_{edges}$ & -1 & -1.23 & 0.40 & -1.68 & -1.23 & -0.74 \\
& & $\theta_{gwesp}$ & 0.2 & 0.26 & 0.18 & 0.01 & 0.30 & 0.49 \\
& & $\theta_{2-stars}$ & -0.3 & -0.37 & 0.10 & -0.49 & -0.35 & -0.26 \\[1ex]
\hline\\[-1ex]
\multirow{3}{*}{50 Nodes} & \multirow{3}{*}{FERGM} & $\theta_{edges}$ & -1 & 37.99 & 6.51 & 35.29 & 38.89 & 42.06 \\
& & $\theta_{gwesp}$ & 0.2 & 0.12 & 0.19 & -0.11 & 0.10 & 0.38 \\
& & $\theta_{2-stars}$ & -0.3 & -7.58 & 1.23 & -8.43 & -7.65 & -7.08 \\[1ex]
\hline\\[-1ex]
\hline\\[-1ex]
\multirow{3}{*}{500 Nodes} & \multirow{3}{*}{mERGM} & $\theta_{edges}$ & -1 & -1.22 & 0.21 & -1.84 & -1.41 & -1.29 \\
& & $\theta_{gwesp}$ & 0.2 & 0.19 & 0.05 & 0.14 & 0.19 & 0.24 \\
& & $\theta_{2-stars}$ & -0.3 & -0.26 & 0.02 & -0.26 & -0.25 & -0.23 \\[1ex]
\hline\\[-1ex]
\multirow{3}{*}{500 Nodes} & \multirow{3}{*}{FERGM} & $\theta_{edges}$ & -1 & -5.18 & 0.04 & -5.24 & -5.18 & -5.12 \\
& & $\theta_{gwesp}$ & 0.2 & 0.18 & 0.04 & 0.14 & 0.19 & 0.23 \\
& & $\theta_{2-stars}$ & -0.3 & 0.06 & 0.01 & 0.05 & 0.06 & 0.06 \\[1ex]
\hline\\[-1ex]
\end{tabular}
\caption{\label{tab:summary_table_mergm_fergm} Resulting estimated means, standard deviations, the medians, 0.1 and 0.9 quantiles of the parameters for network setting with nodal heterogeneity $\sigma_{u}^{2}=1$.}
\end{table}
\FloatBarrier

\section{Examples}
\label{sec:examples}

\subsection{Facebook Network}
\begin{figure}[h]\centering
\includegraphics[scale=0.28]{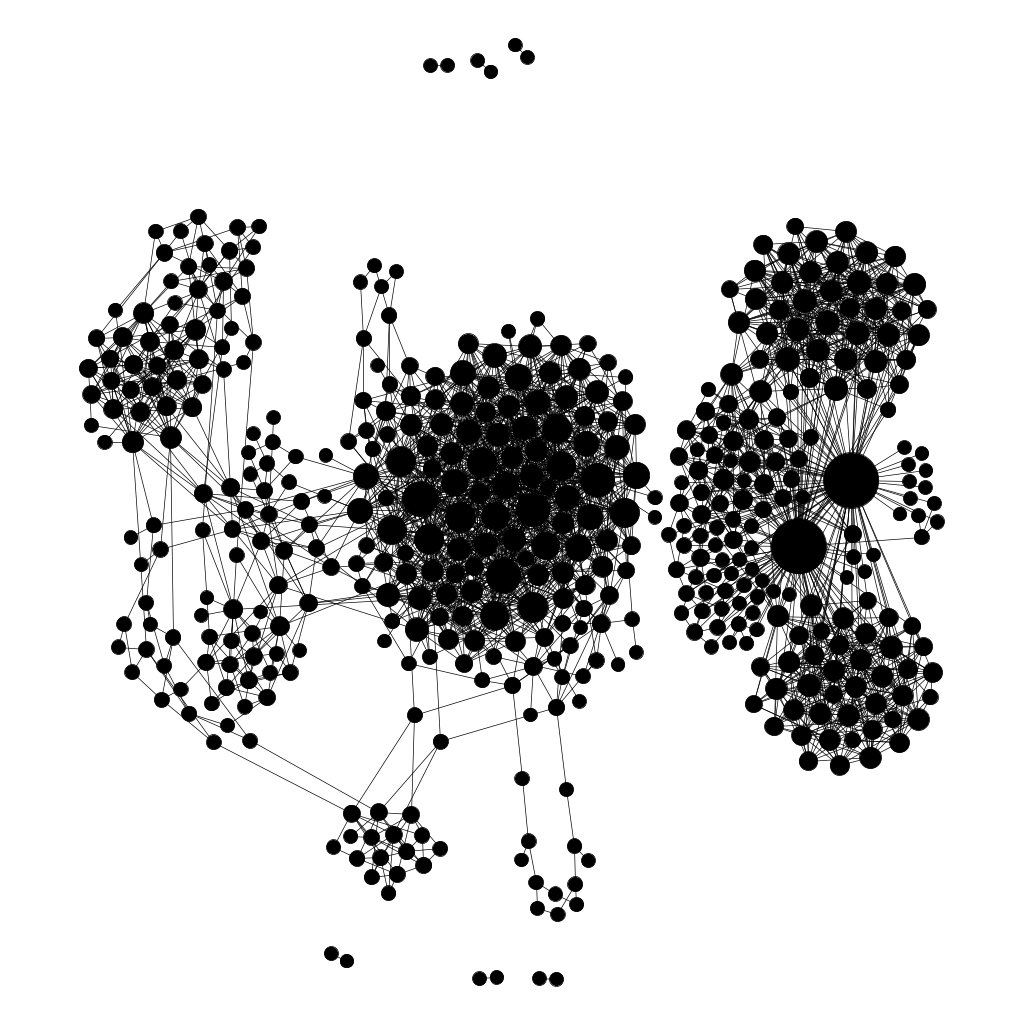} 
\caption{Facebook Network Data. Large nodes indicate nodes with high degrees.}
\label{fig:facebook_network}
\end{figure}

As a first data example, we look at Facebook (undirected) network data, which is publicly accessible \citep{leskovec2012learning}. The entire network comprises 4039 nodes. We analyze a subset of 250 nodes to demonstrate the performance of our routine, e.g. we take the first 251 nodes of the network and remove the "center" node (the "ego"). Figure \ref{fig:facebook_network} gives a visual impression of the network. Just by looking at the network we can easily conclude that an ERGM which assumes nodal homogeneity is more than questionable. The mERGM, therefore, appears as a possible alternative. The aim of our analysis is to evaluate and compare the two models: the mERGM and the standard ERGM, with the intention to quantify the evidence for the presumed favour of the mERGM.\\
We fitted four models to the data, two mERGMs and two ERGMs. Table \ref{tab:netstats} describes the models by listing the sufficient network statistics. As network statistics, we included the number of edges, the number of two-stars and two weighted statistics, i.e. geometrically weighted edgewise shared partners (gwesp) and geometrically weighted nonedgewise shared partners (gwnsp). For the exact definitions of the weighted statistics, we refer to \cite{Snijders-etal:2006}. The number of iterations for the mERGMs was set at 50 to ensure convergence.\\

\FloatBarrier %

\begin{table}[htb!]\centering
\small
\begin{tabular}{lcccccc}
\hline\\[-1ex]
Model type & Model & $\theta_{edges}$ & $\theta_{gwesp}$ & $\theta_{2-stars}$ & $\theta_{gwnsp}$ & Nodal Effects \\[1ex]
\hline\hline\\[-1ex]
\multirow{2}{*}{ERGM} & 1 & \checkmark & \checkmark & \checkmark &  &  \\
 & 2 & \checkmark & \checkmark &  & \checkmark &  \\
\hline\\[-1ex]
\multirow{2}{*}{mERGM} & 3 & \checkmark & \checkmark & \checkmark &  & \checkmark \\
 & 4 & \checkmark & \checkmark &  & \checkmark & \checkmark \\[1ex]
\hline\\[-1ex]
\end{tabular}
\caption{\label{tab:netstats} Models with sufficient network statistics for the Facebook network data.}
\end{table}

\FloatBarrier 
Table \ref{tab:comparison} shows the resulting estimates for the models. Note that the standard deviations given for the mERGM estimates result from the ERGM fit taking the random effects as fixed. As noted above, these estimates are not reliable as they ignore the uncertainty of the estimated random effects. We, therefore, give these values for completeness only but do not interpret them. We see that the gwesp coefficient is always positive, indicating that the probability for an edge between two partners increases with the number of shared partners for the considered edge. The effect is however generally smaller in the mERGM, that is, if node specific heterogeneity is taken into account. To make the models comparable we calculated the AIC values for both the ERGMs and the mERGMs according to the proposed approach as described in Section \ref{sec:selection}. For the calculation of the AIC values, we used 1000 simulations for both ERGMs and mERGMs, respectively. 

\FloatBarrier 
\begin{table}[htb!]\centering
\footnotesize 
\begin{tabular}{lcllll}
\hline\\[-1ex]
Model type & Model & Parameter & Estimate & SE & AIC \\[1ex]
\hline\hline\\[-2ex]%
 \multirow{3}{*}{ERGM} & & $\theta_{edges}$ & -7.178 & 0.091 &  \\
  & 1& $\theta_{gwesp}$ & 1.875 & 0.046 & 6049.086 \\
  &  & $\theta_{2-stars}$ & 0.052 & 0.094 &  \\
\hline\\[-2ex]%
\multirow{3}{*}{ERGM}&  & $\theta_{edges}$ & -5.973 & 1.698 & \\
 & 2& $\theta_{gwesp}$ & 2.076 & 0.881 & 70977.08 \\
 & & $\theta_{gwnsp}$ & -0.034 & 1.211 & \\
\hline\\[-1ex]

\hline\\[-1ex]

 \multirow{3}{*}{mERGM}& & $\theta_{edges}$ & -6.021 & \textit{0.468} & \\
  & 3& $\theta_{gwesp}$ & 1.186 & \textit{0.192} & 4260.723 \\
  & & $\theta_{2-stars}$ & -0.008 & \textit{0.001} &  \\
\hline\\[-2ex]
 \multirow{3}{*}{mERGM}& & $\theta_{edges}$ & -3.943 & \textit{0.698} & \\
 & 4& $\theta_{gwesp}$ & 0.486 & \textit{0.305} & 2867.874 \\
 & & $\theta_{gwnsp}$ & -0.053 & \textit{0.008} & \\[1ex]
\hline\\[-1ex]
\end{tabular}
\caption{\label{tab:comparison} Model fitting results for the Facebook network data. Standard errors listed in the mERGM are not accurate since they ignore the variability resulting through node heterogeneity.}
\end{table}

\begin{figure}[htb!]\centering
  \includegraphics[scale=0.4]{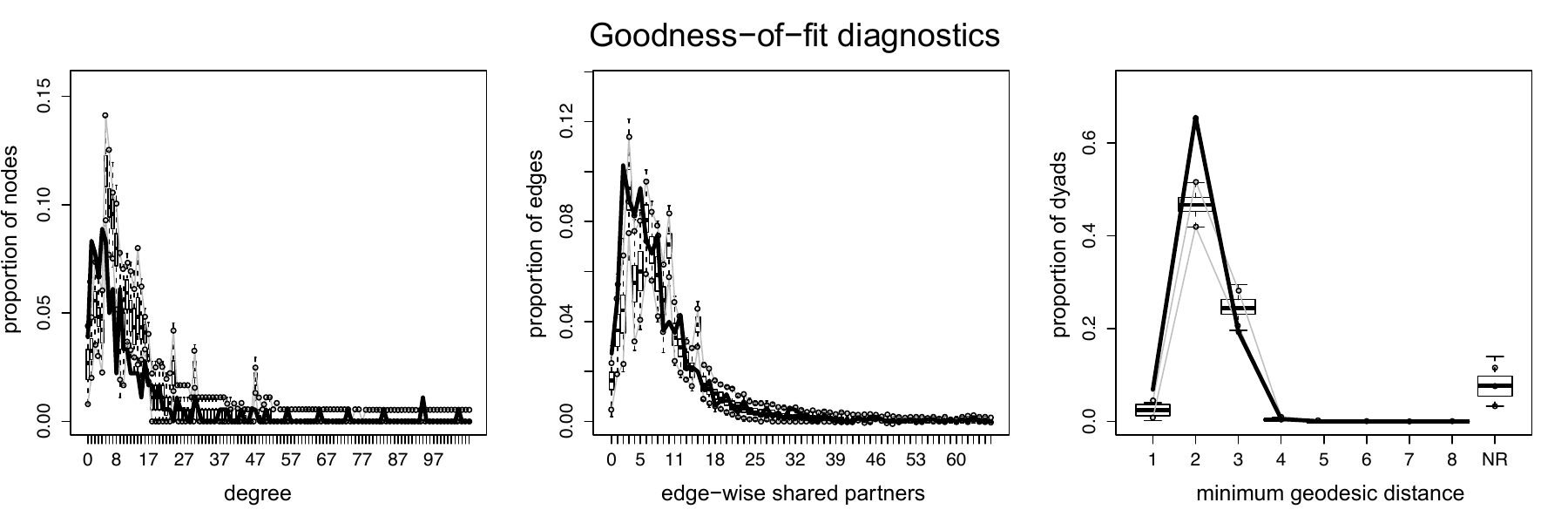} 
  \includegraphics[scale=0.4]{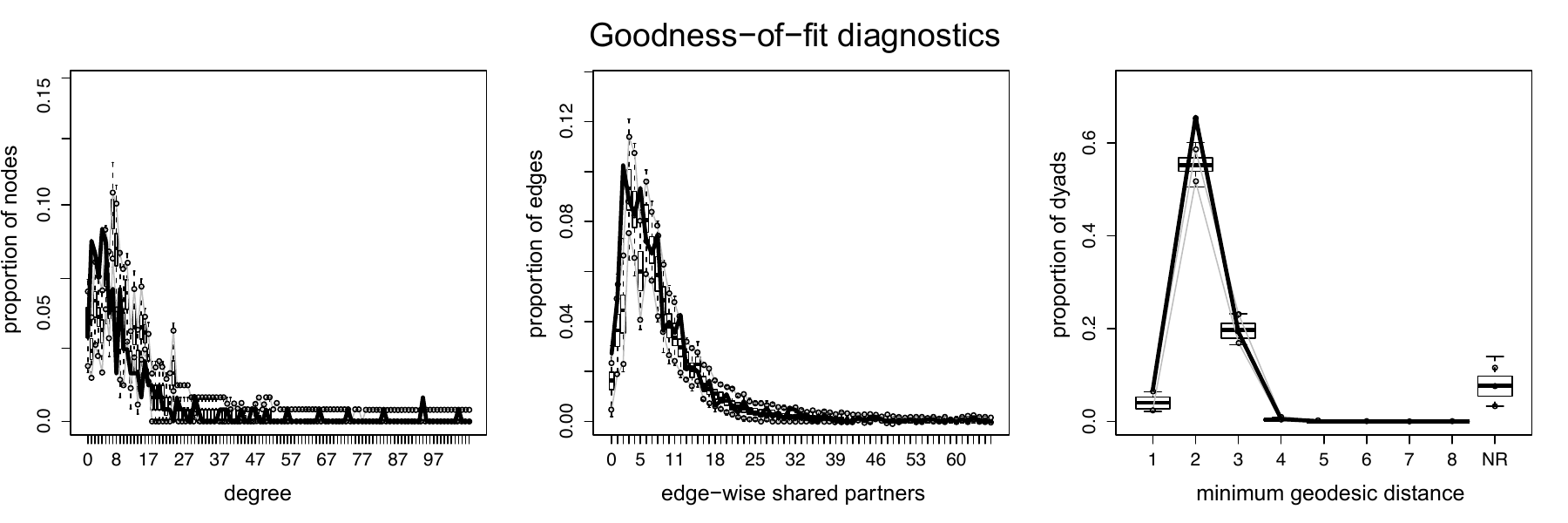} 
\caption{Goodness-of-fit diagnostics for model 1 fitted with ERGM (top row) and for model 3 fitted with mERGM (bottom row) for the Facebook network data.}
\label{fig:fb_model1_model3_gof}
\end{figure}

\begin{figure}[htb!]\centering
  \includegraphics[scale=0.4]{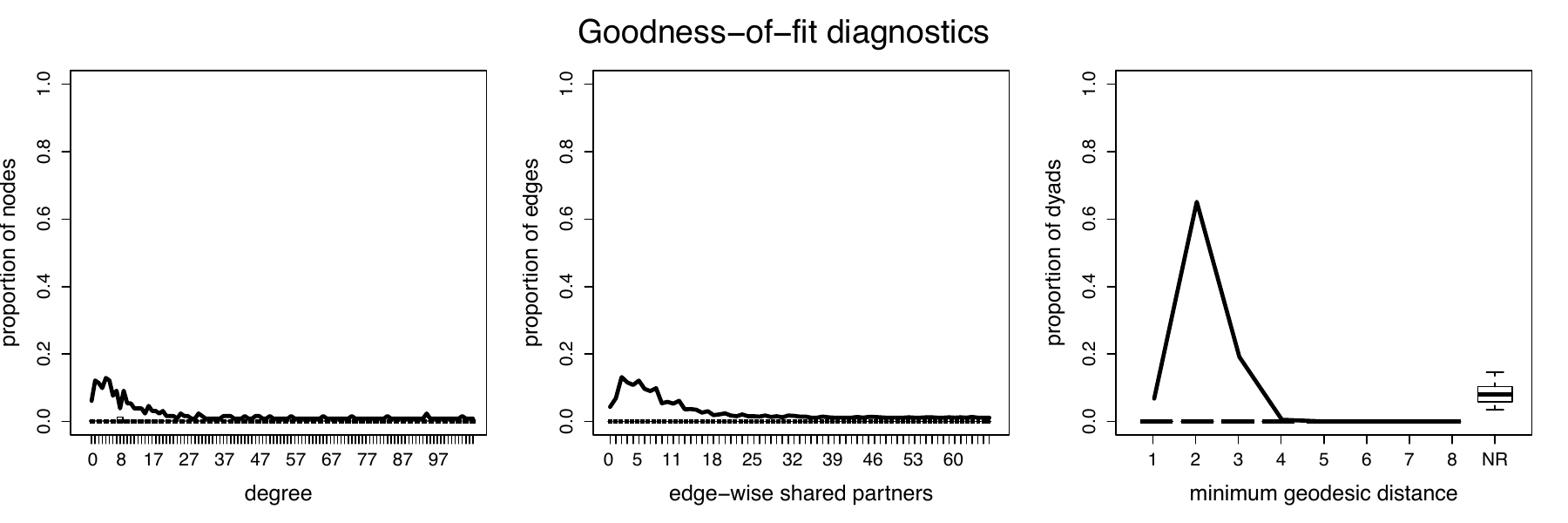} 
  \includegraphics[scale=0.4]{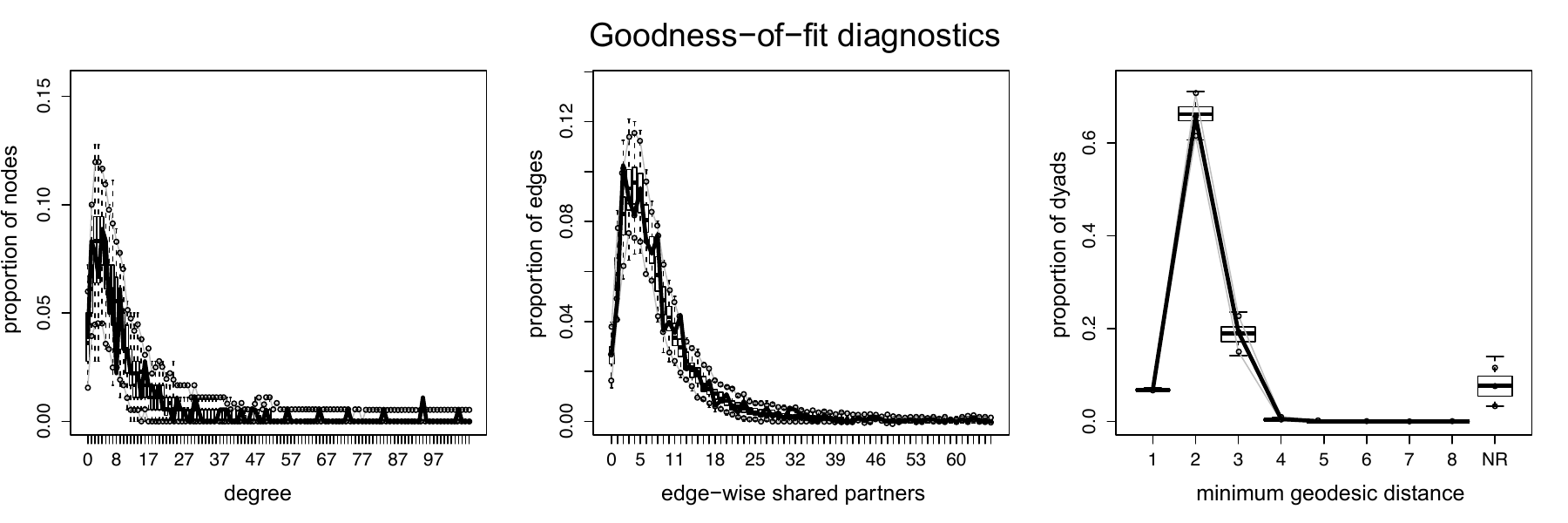} 
\caption{Goodness-of-fit diagnostics for model 2 fitted with ERGM (top row) and for model 4 fitted with mERGM (bottom row) for the Facebook network data.}
\label{fig:fb_model2_model4_gof}
\end{figure}

\FloatBarrier 
Looking at the AIC values of the four models in Table \ref{tab:comparison} we see that both mERGMs outperform the ERGMs. This gives clear evidence of existing node specific heterogeneity, and hence the proposed models with nodal random effects are preferable. This is also apparent in the goodness-of-fit plots. For instance, in Figure \ref{fig:fb_model1_model3_gof} we can see that model 3 fitted with mERGM gives a better fit compared to model 1 fitted with ERGM, which is in line with the corresponding AIC values. Furthermore, the AIC value of model 2 in Table \ref{tab:comparison} fitted with ERGM is exceptionally huge compared to model 4 fitted with mERGM, and we observe this issue well reflected in the goodness-of-fit plots in Figure \ref{fig:fb_model2_model4_gof}, where model 2 struggles with huge convergence issues. Overall, model 4 appears to be the most suitable among the four fitted models to describe the data. 

{}
\FloatBarrier 

\subsection{Zachary's Karate Club Network}
\begin{figure}[htb!]\centering
\includegraphics[scale = 0.45]{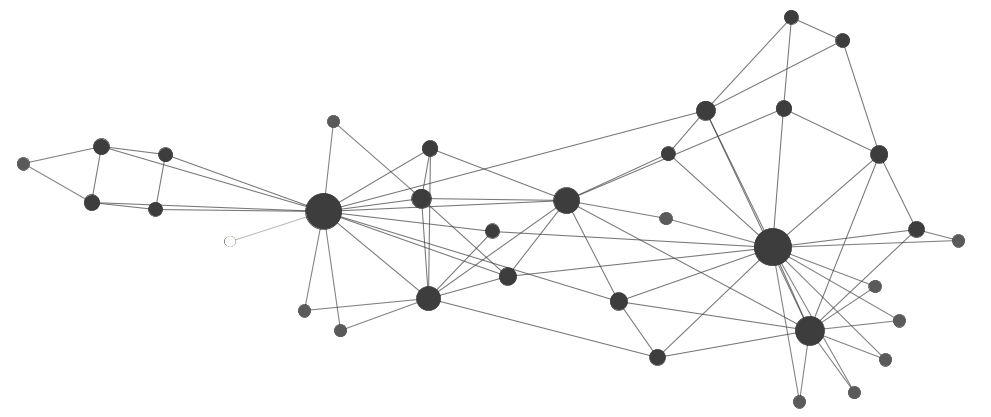} 
\caption{Zachary's Karate Club Network Data. Large nodes indicate nodes with high degrees.}
\label{fig:zach_net}
\end{figure}

As a second data example, we look at a well known dataset, the Zachary's karate club \citep{Zachary:1977}. This undirected network data represents the friendship among 34 members of a university karate club. Figure \ref{fig:zach_net} shows the network graph of Zachary's Karate Club. One can easily see that in this network, there are few nodes with high degree, while the remaining nodes have only few edges, so again, we assume that the mERGM should be a suitable approach to capture the unobserved nodal-heterogeneity in the network. We fitted three different ERGMs and three mERGMs to this data. To make the models comparable, we included the same network statistics. Table \ref{tab:netstats_zach} gives an overview of the different models. In Table \ref{tab:estimation_zach} we summarize the results of our models including the AIC values. The iteration steps for the mERGMs was set to 50. For the calculation of the AIC values, we used 1000 simulations for both ERGMs and mERGMs.

\begin{figure}[htb!]\centering
  \includegraphics[scale=0.46]{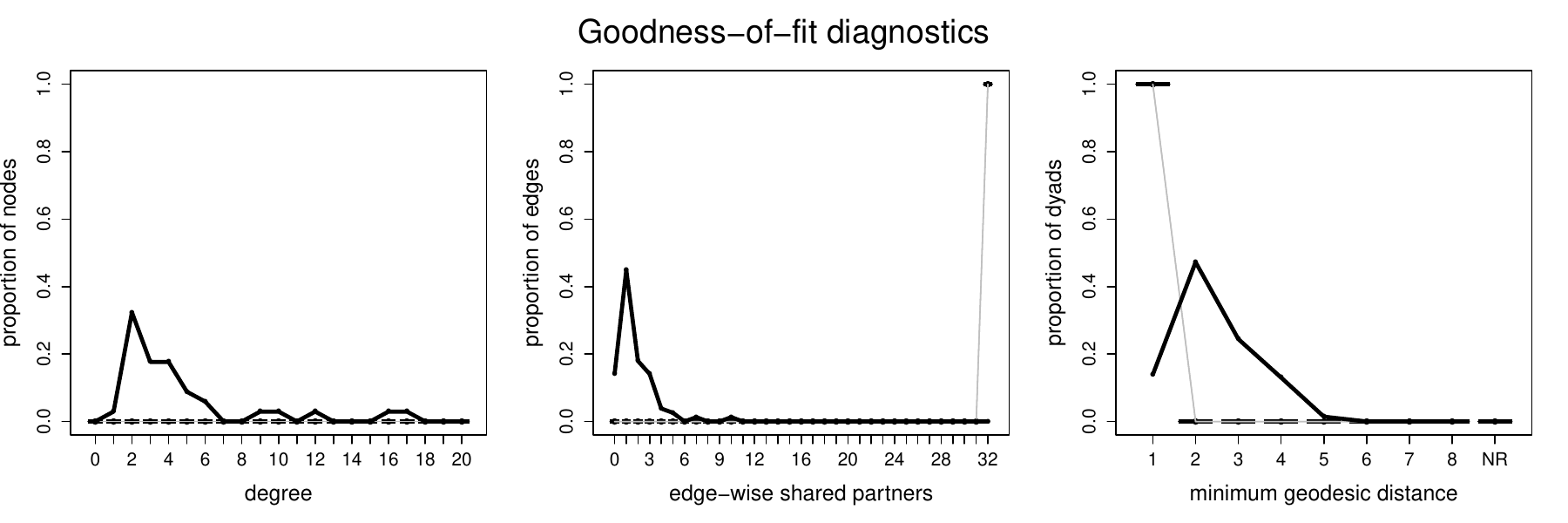} 
  \includegraphics[scale=0.46]{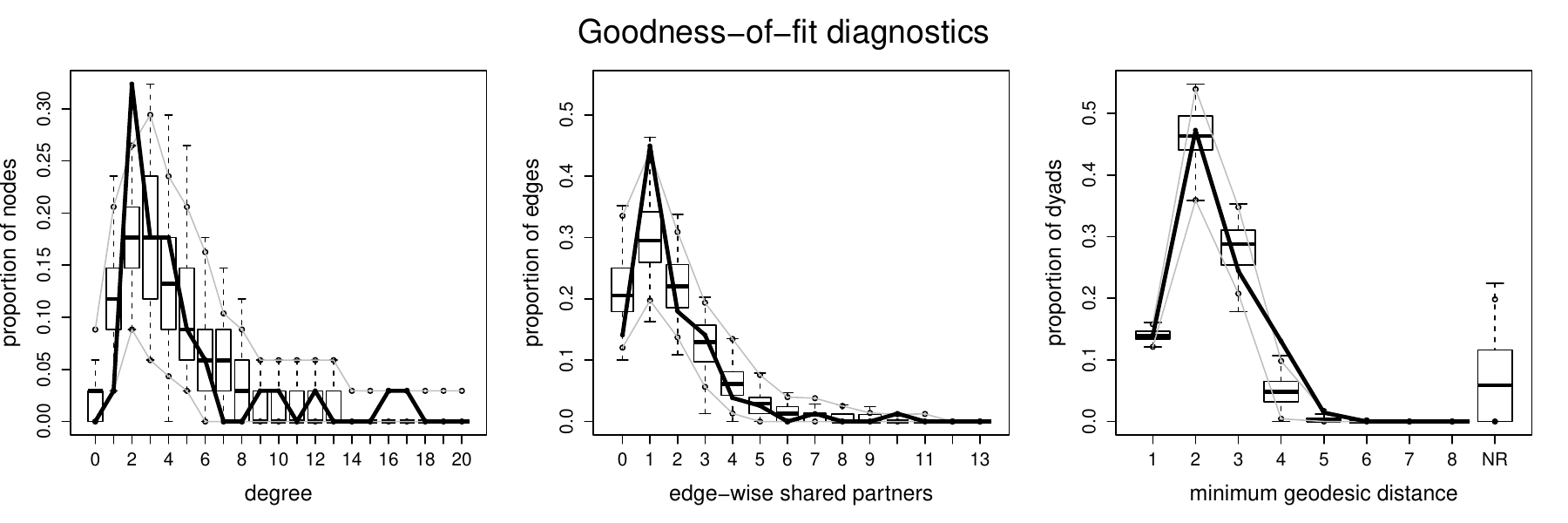} 
\caption{Goodness-of-fit diagnostics for model 1 fitted with ERGM (top row) and for model 4 fitted with mERGM (bottom row) for the karate club data.}
\label{fig:zach_model1_model4_gof}
\end{figure}

We can see that model 1 fitted with ERGM struggles with convergence issues. This is mirrored in invalid variance estimates, resulting from a badly conditioned Fisher matrix. We, therefore, indicate this as $"\star"$ in Table \ref{tab:estimation_zach}, which also means, of course, that the estimate itself is not reliable at all. We refer to \cite{HunterHandcock-etal:2008} for further explanations. On the other hand, model 4 fitted with mERGM with the same model parameters as model 1 does not show any convergence issues, which also means that the mERGM can deal with estimation degeneracy issues. The inclusion of node specific heterogeneity works towards numerical stabilization. To explore this in more depth, we look in Figure \ref{fig:zach_model1_model4_gof} at the goodness-of-fit plots for model 1 fitted with ERGM. Figure \ref{fig:zach_model1_model4_gof} shows the same diagnostics results for model 4 fitted with mERGM. Remember that these two models include the same sufficient network statistics. Boxplots of the distributions of degree, edge-wise shared partners and minimum geodesic distance for the resulting simulated networks are shown in the plots where the bold line indicates the values of the original karate club dataset. 
In the diagnostics plots of model 1 in Figure \ref{fig:zach_model1_model4_gof} we can clearly see that ERGM fails to fit the model, whereas the diagnostics plots of model 4 in Figure \ref{fig:zach_model1_model4_gof} gives a good evidence of an appropriate fit.

\begin{table}[htb!]\centering
\footnotesize 
\begin{tabular}{lccccccc}
\hline\\[-1ex]
Model type & Model & $\theta_{edges}$ & $\theta_{gwesp}$ & $\theta_{2-stars}$ & $\theta_{gwnsp}$ & $\theta_{gwdegree}$ & Nodal Effects \\[1ex]
\hline\hline\\[-1ex]
\multirow{3}{*}{ERGM} & 1 & \checkmark & \checkmark & \checkmark &  &  & \\
  & 2& \checkmark & \checkmark &  & \checkmark &  & \\
  & 3& \checkmark & \checkmark &  &  & \checkmark & \\
\hline\\[-1ex]
\multirow{3}{*}{mERGM} & 4 & \checkmark & \checkmark & \checkmark &  & & \checkmark \\
  & 5& \checkmark & \checkmark &  & \checkmark  &  & \checkmark \\
  & 6& \checkmark & \checkmark &  &  & \checkmark & \checkmark \\[1ex]
\hline\\[-1ex]
\end{tabular}
\caption{\label{tab:netstats_zach} Models with sufficient network statistics for the Zachary network data.}
\end{table}

\begin{table}[htb!]\centering
\footnotesize 
\begin{tabular}{lcllll}
\hline\\[-1ex]
Model type & Model & Parameter & Estimate & SE & AIC \\[1ex]
\hline\hline\\[-2ex]
 \multirow{3}{*}{ERGM}&  & $\theta_{edges}$ & -4.893 & $\star \star \star$ &  \\
 & 1& $\theta_{gwesp}$ & 0.642 & $\star \star \star$ & $\star \star \star$ \\
 & & $\theta_{2-stars}$ & 0.689 & $\star \star \star$ & \\[1ex]
 \hline\\[-2ex]
 \multirow{3}{*}{ERGM}& & $\theta_{edges}$ & -3.635 & 0.241 &  \\
 & 2& $\theta_{gwesp}$  & 0.596 & 0.117 & 496.351 \\
 & & $\theta_{gwnsp}$ & 0.153 & 0.026 &  \\[1ex]
\hline\\[-2ex]
\multirow{3}{*}{ERGM}& & $\theta_{edges}$ & -3.183 & 0.513 & \\
 & 3& $\theta_{gwesp}$ & 0.716 & 0.181 & 442.883 \\
 & & $\theta_{gwdegree}$ & -0.519 & 0.8990 & \\[1ex]
\hline\\[-1ex]

\hline\\[-1ex]

\hline\\[-1ex]

\multirow{3}{*}{mERGM}&  & $\theta_{edges}$ & -1.214 & \textit{0.386} & \\
 & 4& $\theta_{gwesp}$ & 0.236 & \textit{0.174} & 311.304 \\
 & & $\theta_{2-stars}$ & -0.159 & \textit{0.042} &  \\[1ex]
 \hline\\[-2ex]
\multirow{3}{*}{mERGM}& & $\theta_{edges}$ & -1.776 & \textit{0.366} &  \\
 & 5& $\theta_{gwesp}$  & -0.144 & \textit{0.109} &  337.894\\
 & & $\theta_{gwnsp}$ & -0.089 &  \textit{0.043} & \\[1ex]
\hline\\[-2ex]
 \multirow{3}{*}{mERGM}&  & $\theta_{edges}$ & -4.464 & \textit{0.724} & \\
 & 6& $\theta_{gwesp}$ & 0.213 & \textit{0.176} & 303.217 \\
 & & $\theta_{gwdegree}$ & 4.427 & \textit{1.341} & \\[1ex]
\hline\\[-1ex]
\end{tabular}
\caption{\label{tab:estimation_zach} Model fitting results for the Zachary network data. Standard errors listed in the mERGM are not accurate since they ignore the variability resulting through node heterogeneity.}
\end{table}

\begin{figure}[htb!]\centering
  \includegraphics[scale=0.46]{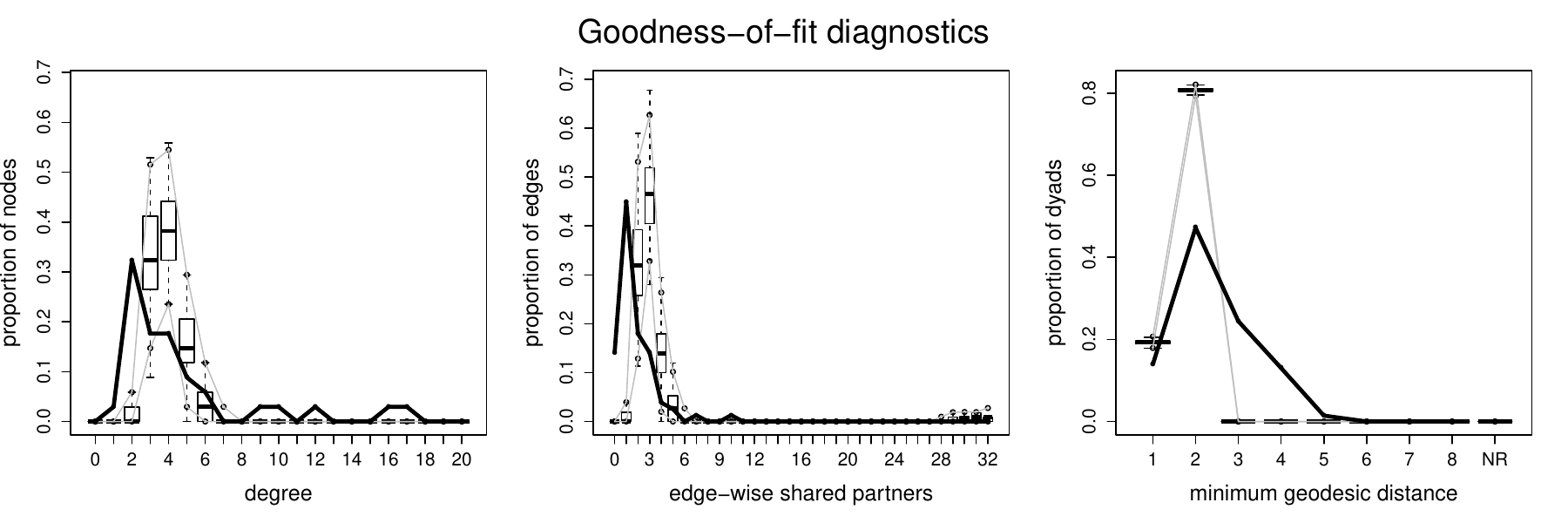} 
  \includegraphics[scale=0.46]{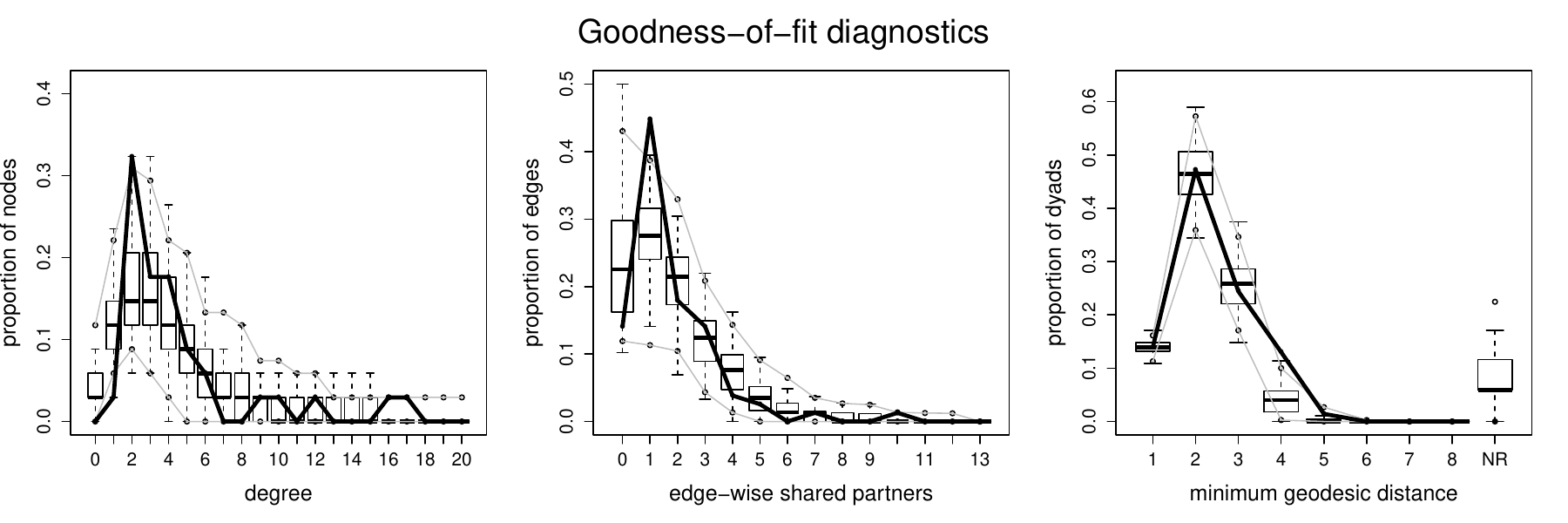} 
\caption{Goodness-of-fit diagnostics for model 2 fitted with ERGM (top row) and for model 5 fitted with mERGM (bottom row) for the karate club data.}
\label{fig:zach_model2_model5_gof}
\end{figure}

We extend the model exploration to the other models. Figure \ref{fig:zach_model2_model5_gof} shows the goodness-of-fit diagnostics plots of both model 2 fitted with ERGM and model 5 fitted with mERGM, respectively. For model 2, we see some problems in Figure \ref{fig:zach_model2_model5_gof} concerning all the three diagnostics, the degree distribution, the edgewise-shared partners distribution and the minimum geodesic distance, which indicate the poorness of the model. For model 4, in contrast, we can see in Figure \ref{fig:zach_model2_model5_gof} again a much better performance of the fit. 
\FloatBarrier

\begin{figure}[htb!]\centering
  \includegraphics[scale=0.46]{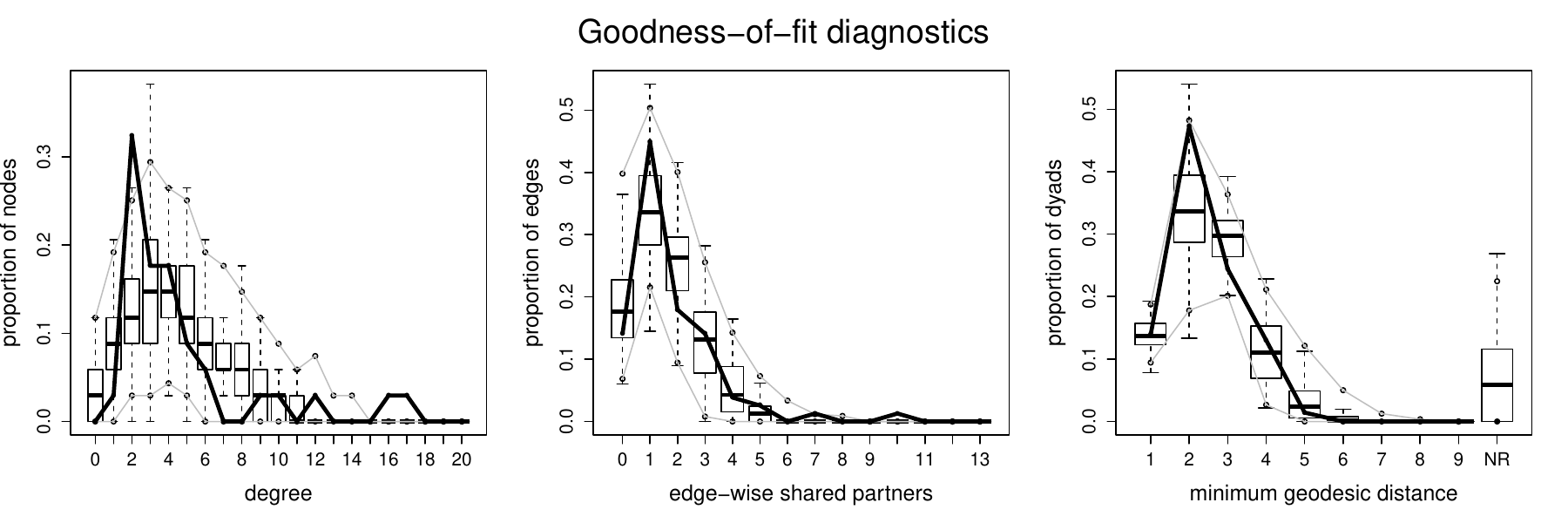}
  \includegraphics[scale=0.46]{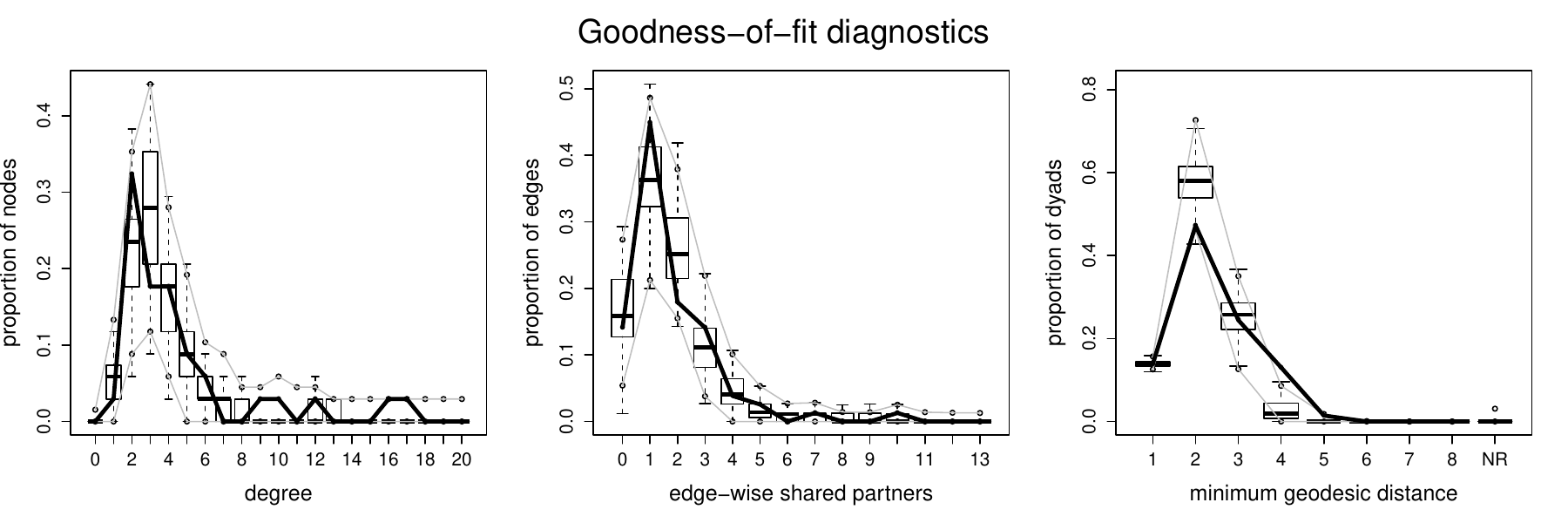} 
\caption{Goodness-of-fit diagnostics for model 3 fitted with ERGM (top row) and for model 6 fitted with mERGM (bottom row) for the karate club data.}
\label{fig:zach_model3_model6_gof}
\end{figure}

Finally, Figure \ref{fig:zach_model3_model6_gof} shows the diagnostics plots of model 3 fitted with the ERGM, this model is the best ERGM fitted to this data according to the AIC value and also the diagnostics plots are reasonable. On the other hand, model 6 fitted with mERGM, including the same sufficient network statistics as model 3, is the best mERGM fitted to this data according to the AIC value. However, the goodness-of-fit of model 6 shown in Figure \ref{fig:zach_model3_model6_gof} visually looks better than of model 3, which also justifies with a smaller AIC value.

\subsection{High School Friendship Network}
As a third data example, we investigate a real-world dataset, a high school friendship network in Marseilles, France, provided by \cite{mastrandrea2015contact}. This network data represents the friendship among 134 high school students of specific classes. These specific classes, unique to the French educational system, gather students for studies that last two years after completing the usual high school studies. These classes are in a different part of the building, so the students are somehow isolated from the "regular" high school students. As a result, they form an almost closed population with little contact with the outside world, at least during schooldays. At the end of these two years, these students take competitive exams to gain admission to various higher educational institutions. The classes have different specializations: "MP" classes focus more on mathematics and physics, "PC" classes on physics and chemistry, "PSI" classes on engineering, and "BIO" classes on biology. Furthermore, there are three classes of type "MP", two of type "PC", one of type "PSI" and three of type "BIO". Due to the class sizes in the dataset, as demonstrated in Table \ref{tab:no_students}, we decided to merge the types of classes to get an appropriate fit for our models, e.g., we do not distinguish between the types of the classes. 

\begin{table}[htb!]\centering
\footnotesize 
\begin{tabular}{|l|c|c|c|c|c|c|c|c|c|}
\hline
Classes & MP & MP*1 & MP*2 & PC & PC* & PSI* & 2BIO1 & 2BIO2 & 2BIO3 \\ \hline 
No. of Students & 21 & 3 & 7 & 21 & 10 & 15 & 10 & 19 & 28 \\
\hline
\end{tabular}
\caption{\label{tab:no_students} Number of students according to the types of the classes in the data.}
\end{table}

\begin{figure}[htb!]\centering
\includegraphics[scale = 0.42]{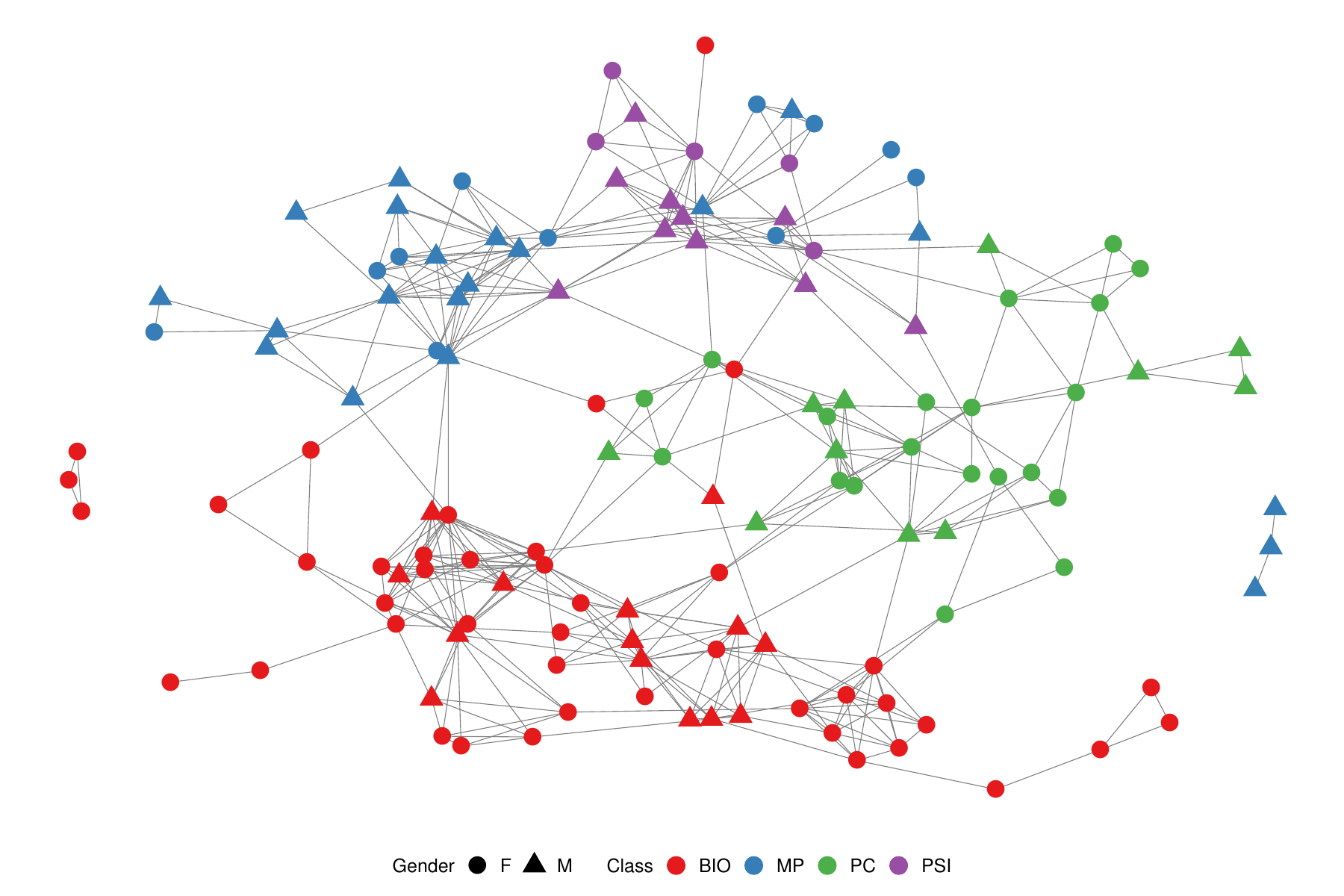} 
\caption{High School Friendship Network Data. The Colour of the nodes represents the different classes. The shape of the nodes describes the gender of the students.}
\label{fig:school_friend_net}
\end{figure}

 Additionally, we also have the gender information of the students in the data, with 79 female and 55 male students. Figure \ref{fig:school_friend_net} shows the friendship network graph, the colour of the nodes represents the different classes, where the shape of the nodes the gender of the students. Assuming that there might be unobserved nodal heterogeneity, which can not be captured exclusively by the nodal covariates, the mERGM could be a reasonable approach.

\begin{figure}[htb!]\centering
  \includegraphics[scale=0.4]{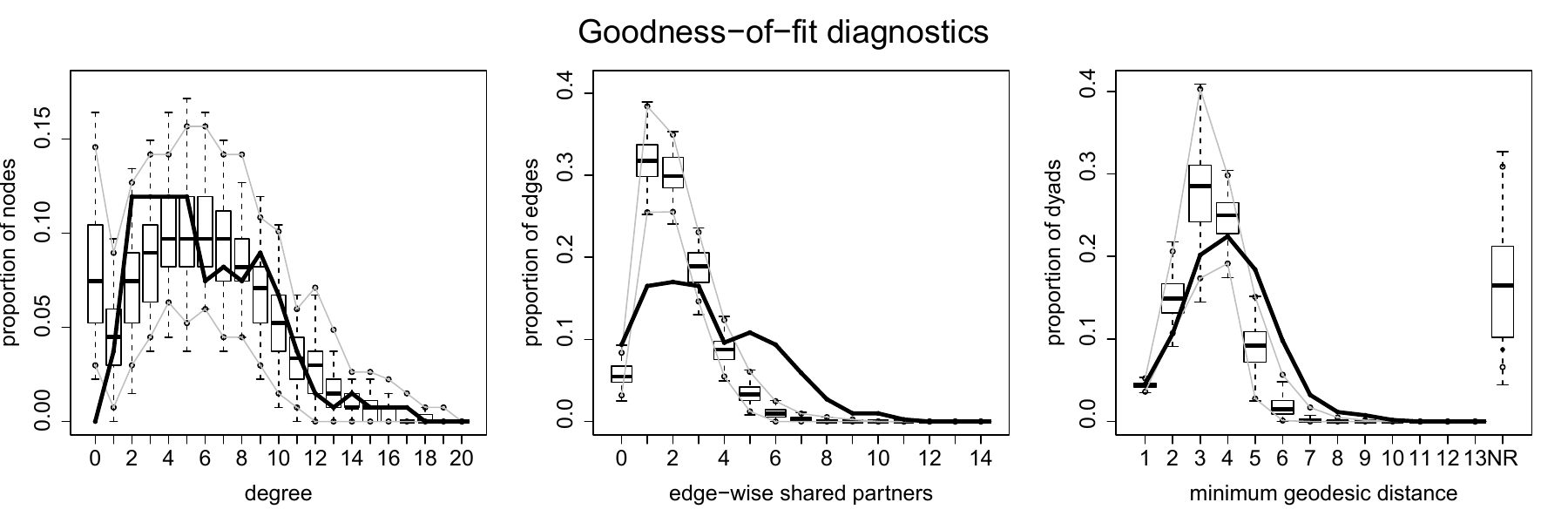}
  \includegraphics[scale=0.4]{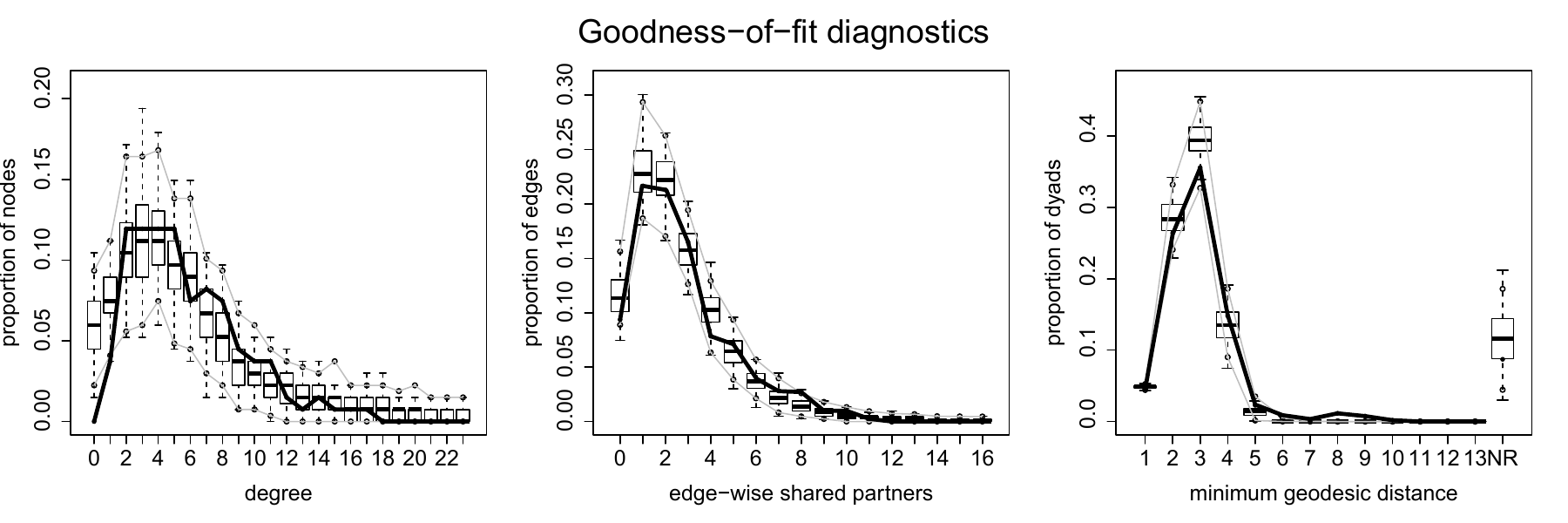} 
\caption{Goodness-of-fit diagnostics for the model fitted with ERGM (top row) and for the model fitted with mERGM (bottom row) for the high school friendship network data.}
\label{fig:friend_ergm_mergm_gof}
\end{figure}

We fit an ERGM and a mERGM to this data, including dyad-dependent network statistics such as the \texttt{GWESP} (geometrically weighted edgewise shared partner distribution) with a fixed decay parameter equal to $0.25$ and dyad-independent terms such as the \texttt{nodefactor} and \texttt{nodematch} parameters. The \texttt{nodematch} is the homophily parameter in ERGM, where we also allow for each class to have a unique propensity for within-class ties, we refer to \cite{morris2008specification} for instance, for more details regarding the ERGM terms and their specifications. The iteration steps for the mERGMs was set to 50. For the calculation of the AIC values, we used 1000 simulations for both ERGMs and mERGMs. In Table \ref{tab:estimation_school_friendship} we summarize the results of the fitted models. Comparing the AIC values of the two models, the model fitted with mERGM is preferred, which is not very surprising since the mERGM additionally takes the "unobserved" nodal-heterogeneity into account, which is not captured by dyad-dependent or independent terms, aka network statistics. For the sake of interpretation, we look, for instance, at the log-odds of a hypothetical tie between two male students attending the MP class that does not close a triangle. In ERGM the log-odds of such a tie would be $-4.264$; in mERGM, the interpretation is slightly different since it should be only in the conditional sense. This means, given two male students that attend the MP class, the conditional log-odds of such a tie is $-1.671$. In Figure \ref{fig:friend_ergm_mergm_gof} we show the diagnostic plots of the two models, we can clearly see that the model fitted with the mERGM performs much better, which is also clear evidence to the corresponding AIC values. 

\begin{table}[htb!]\centering
\footnotesize
\begin{tabular}{llllllll}
\hline\\[-1ex]
Model type & Parameter & Estimate & SE  & Model type & Parameter & Estimate & SE  \\[1ex]
\hline\hline\\[-2ex]
\multirow{11}{*}{ERGM} & Edges & -7.606 & 0.403 & \multirow{11}{*}{mERGM} & Edges & -9.021 & \textit{0.415}  \\
                      & GWESP (0.25) & 2.026 & 0.165 & & GWESP (0.25) & 0.538 & \textit{0.215} \\
                      & Male & 0.157 & 0.052 & &Male & -0.494 & \textit{0.068} \\
                      & Gender match & 0.226 & 0.089 & &Gender match & 0.813 & \textit{0.099} \\
                      & MP & 0.627 & 0.272 & &MP & 2.355 & \textit{0.238} \\
                      & PC & 0.309 & 0.272 & &PC & 1.253 & \textit{0.238} \\
                      & PSI & 1.054 & 0.276 & &PSI & 3.042 & \textit{0.221} \\
                      & BIO class match & 2.562 & 0.402 & &BIO class match & 4.403 & \textit{0.295} \\
                      & MP class match & 1.547 & 0.337 & &MP class match & 2.814 & \textit{0.349} \\
                      & PC class match & 2.279 & 0.393 & &PC class match & 3.572 & \textit{0.379} \\
                      & PSI class match & 1.504 & 0.379 & &PSI class match & 2.047 & \textit{0.355} \\[1ex]
\hline\\[-1ex]
AIC & 3880.681 & & & AIC & 1097.836 &  &  \\[1ex]
\hline\\[-1ex]
\end{tabular}
\caption{\label{tab:estimation_school_friendship} Model estimates for the high school friendship network. Standard errors listed in the mERGM are not accurate since they ignore the variability resulting through node heterogeneity.}
\end{table}

\FloatBarrier

\section{Discussion}
\label{sec:discussion}

In most cases, nodal heterogeneity in the network is explained by including known or well-studied nodal covariates, see e.g. \cite{Robins-etal:2001}. However, the node-specific covariates cannot fully or sufficiently account for unobserved heterogeneity in the network. Our extensions towards Mixed Exponential Random Graph Models can therefore be a meaningful approach to model network data by just adding nodal random effects to the model to capture the unobserved nodal heterogeneity.\\

Our proposed model (\ref{eq:logextergm}) can in principle also be extended to directed networks, where $u^{(s)}_{i}$ and $u^{(r)}_{j}$ would be treated as random sender and random receiver node specific coefficients, respectively. We consider it beyond the scope of the current paper. \\

Though the calculation of the AIC value is computationally intensive, our proposed method of estimating and calculating the AIC values allows us to compare the mERGM with the conventional ERGM. Furthermore, as we can see in our simulation study, the mERGM can always be a reasonable approach for modelling networks even if we observe small nodal heterogeneity in the network. Overall, the mERGM works towards stabilizing the fitting routine without adding too much numerical effort.\\

In this manuscript, neither in the simulation study nor in the examples, we investigate assortative mixing parameters. Therefore, a more thorough investigation in this regard is worth pursuing in future work.

\section*{Acknowledgements}
\label{sec:acknowledgements}
The project was supported by the European Cooperation in Science and Technology [COST Action CA15109 (COSTNET)].

\section*{Fundings}
\label{sec:funding}
This work has been funded by the German Federal Ministry of Education and Research (BMBF) under Grant No. 01IS18036A. The authors of this work take full responsibility for its content.

\clearpage
\bibliography{literatur}

\end{document}